\documentclass[12pt,journal,onecolumn]{IEEEtran}
\usepackage[margin=1in]{geometry}
\usepackage{times}
\usepackage{color}
\usepackage{verbatim}

\usepackage{cite}
\usepackage{amsmath}
\usepackage{amssymb}
\usepackage{sgame}
\usepackage{algorithmic}%
\usepackage{algorithm}%

\usepackage{graphicx}
\newcommand{\theset}[1]{\left\{#1\right\}}

\newcommand{\magn}[1]{\left\vert #1 \right\vert}
\newcommand{\st}{~\vert~}

\newtheorem{proposition}{Proposition}
\newtheorem{theorem}{Theorem}
\newtheorem{claim}{Claim}

\newtheorem{definition}{Definition}

\newenvironment{proof}{\textbf{Proof.}}{\QED}

\begin{document}
\baselineskip 6.5mm
\title{BLMA: A Blind Matching Algorithm with Application to Cognitive Radio Networks}
%\author{Doha {Hamza}\thanks{doha.hamzamohamed@kaust.edu.sa} \and Jeff S. Shamma\thanks{jeff.shamma@kaust.edu.sa}\\ 
%\small King Abdullah University of Science and Technology (KAUST)\\
%\small Computer, Electrical, and Mathematical Sciences and Engineering  Division (CEMSE)\\
%\small Thuwal, Saudi Arabia}

%\author{
 %   \IEEEauthorblockN{Doha Hamza, and Jeff S. Shamma}\\
 %   \IEEEauthorblockA{\small Computer, Electrical, and Mathematical Science and Engineering Division (CEMSE)\\
%King Abdullah University of Science and Technology (KAUST),
%Thuwal, Saudi Arabia
 %   \\\small Emails: {\{doha.hamzamohamed, jeff.shamma\}@kaust.edu.sa}
%}}

\author{
\IEEEauthorblockN{Doha Hamza{$^*$}, and Jeff S. Shamma{$^*$,$^\dagger$}}\\
    \IEEEauthorblockA{%
        {$^*$}Computer, Electrical, and Mathematical Science and Engineering (CEMSE) Division, King Abdullah University of Science and Technology (KAUST), Thuwal, Kingdom of Saudi Arabia\\
				${^\dagger}$School of Electrical and Computer Engineering, Georgia Institute of Technology, Atlanta, GA%
				\\Emails: {\{doha.hamzamohamed, jeff.shamma\}@kaust.edu.sa}
    }
    
    }

\maketitle
%\begin{IEEEkeywords}
%Cognitive radio.
%\end{IEEEkeywords}

%\textbf{Keywords:}
%==================================================================================================
%The rest of the paper is organized as follows. The system model is provided in Section \ref{SystemModel}. We formulate the problem in Section %\ref{Problem_Formulation}. Section \ref{numres} contains some numerical results whereas Section \ref{Conclusion} concludes the paper.

%%%%%%%%%%%%%%%%%%%%%%%%%%%%%%%%%%%%%%%%%%%%%%%%%%%%%%%%%%%%%%
\begin{abstract}
{

%\textcolor{red}{$b_\ell+\epsilon$}

We consider a two-sided matching problem with a defined notion of pairwise stability. We propose a distributed blind matching algorithm (BLMA) to solve the problem. We prove the solution produced by BLMA will converge to an $\epsilon$-pairwise stable outcome with probability one. We then consider a matching problem in cognitive radio networks. Secondary users (SUs) are allowed access time to the spectrum belonging to the primary users (PUs) provided that they relay primary messages. We propose a realization of the BLMA to produce an $\epsilon$-pairwise stable solution assuming quasi-convex and quasi-concave utilities. In the case of more general utility forms, we show another BLMA realization to provide a stable solution. Furthermore, we propose negotiation mechanism to bias the algorithm towards one side of the market. We use this mechanism to protect the exclusive rights of the PUs to the spectrum. In all such implementations of the BLMA, we impose a limited information exchange in the network so that agents can only calculate their own utilities, but no information is available about the utilities of any other users in the network.}

\end{abstract}
%%%%%%%%%%%%%%%%%%%%%%%%%%%%%%%%%%%%%%%%%%%%%%%%%%%%%%%%%%%%%%
\section{Introduction}

%The problem at hand fits well within the framework of matching theory since we can divide the sets of agents into two distinct sets with opposing interests.
The assignment game is a model for a two-sided matching market with transferable linear utilities\cite{shsh}. Shapley and Shubik first formulated this game in coalitional form in {a housing market}. The outcome of the game is an assignment of prices to homes together with a feasible allocation of homes to buyers (a matching) \cite{alaei}. In this (linear) assignment game model, utility transfers occur at a one-to-one rate from one player to any other \cite{serr}. In the case of general nonlinear utilities, different agents' utilities are not measured in the same units. This induces another class of games called the generalized assignment {game}\cite{roth_book}.

An important solution concept in such games is pairwise stability. Pairwise stable vectors are payoff vectors that cannot be improved upon by any pair of players. Shapley and Shubik showed that there always exists pairwise stable solutions in the assignment game and they coincide with the core of the game\cite{np}.

In this paper, we begin by formulating a context-free matching problem\footnote{As argued in \cite{np}, the main difference between marriage problems (or matching problems) and assignment games is that agents in the former have ordinal preferences over agents on the opposite side of the market and thus this game is said to be a nontransferable utility (NTU) cooperative game, while in linear assignment games, on the other hand, agents have cardinal utilities which they can transfer to agents on the other side of the market. Such games are called transferable utility (TU) cooperative games. Assignment games with non-linear utilities, however, fall under the class of NTU games since different agents' utilities are not measured in the same units and are therefore non-transferable \cite{alaei, dutting}. Bearing all this in mind, we will thereafter refer to our problem, which encompasses both linear and generalized assignment games, as a matching problem.}. In our model, agents maintain aspiration levels which are an abstraction of their potential utility from a match. We provide a mechanism where these aspiration levels go up when agents' search for a match is fruitful and they go down otherwise.  We demonstrate that our model is general enough to encompass the linear and generalized assignment game frameworks. Our solution concept is a modified notion of pairwise stability such that the payoff vectors, the aspiration levels, cannot be $\epsilon$-improved upon by any pair of players. After defining our problem, we propose a blind matching algorithm (BLMA) to produce a matching accompanied by vectors of aspiration levels in the two-sided matching problem. We show that the BLMA will converge to our defined notion of pairwise stability with probability one. 

In its abstract form, the proposed BLMA does not specify how agents negotiate their way into a matching. Neither does the algorithm specify how agents eventually update their aspirations if they match. The algorithm also does not specify how much information agents know about each other's aspiration levels. All that is required for assured convergence is that agents have a mechanism of knowing if they have agreeable matches and if there exists such agreement of two agents' aspirations, that their match occurs with positive probability. It is in that sense that the algorithm is both blind and non-deterministic as the details of all the aforementioned concerns are left to the specific context of the matching problem.

Next, considering the context of cooperative spectrum sharing in cognitive radios, we formulate a matching problem where secondary users (SUs) relay primary user ({PU}) data in exchange for spectrum access time. Agents' utilities are finite but non-linear in the time and power resources. Furthermore, the PU's utility is quasi-convex while the SU's utility is quasi-concave\footnote{Where clear and for convenience, we will thereafter use the term quasi-convex to refer to both the quasi-convex and quasi-concave properties.}. The desired outcome is a matching of the PUs and the SUs and a specification of a set of $\epsilon$-pairwise stable utilities. In this specific context, exploiting the quasi-convexity property, we provide a negotiation mechanism that randomly splits the time and power resources among any two agents. Because such mechanism is only a specific realization of the BLMA, we are assured convergence to a matching with $\epsilon$-pairwise stable utilities. 

We then relax our assumptions further and modify the cooperative relaying mechanism so that agents' utilities are no longer quasi-convex \footnote{In this modified setting, agents on the two sides of the market will still show opposite trends in the time and power resources.}. In this case, we propose a different negotiation mechanism where agents flip a coin to decide whether to optimize the power or time resource while maintaining all other previous agreements. This simple decision process of working one dimension at a time can also be shown to be another specific negotiation mechanism for the BLMA. Hence, we show convergence to pairwise stable solutions despite the lack of quasi-convexity. In all such negotiation mechanisms, matched users update their aspiration levels by randomly selecting a point in their agreement set. 

Finally we make an observation that restricting the time {component} of the agents offers to small values, is an effective mechanism to bias the matching outcome towards one side of the market, the PUs in our case. This provides a simple scheme to protect {PU}'s rights to the wireless spectrum. Since the search space is also significantly reduced, faster convergence comes as a bonus. 

%To summarize this section, the main contributions of our work are as follows:
%\begin{itemize}
%\item {\emph{A Context-free Matching Problem}}: We formulate an abstract matching problem that is shown to encompass both the linear and generalized assignment game models. 
%\item  {\emph{A blind matching algorithm}}: We propose a simple algorithm to pair agents in the matching problem and prove the algorithm converges to $\epsilon$-pairwise stable solutions with probability one. PUs and SUs with pairwise stable power and time allocations. Despite the limited information restriction, the algorithm is shown to converge to core solutions in finite time. 
%\item {\emph{A blind matching algorithm with improved convergence properties}}: We also propose a modified blind matching algorithm with a different match point selection mechanism {that} is shown to improve the convergence of the BLMA algorithm. 
%\item {\emph{Scheme to b

The rest of this paper is organized as follows. {Section \ref{sec:lit} reviews} the related literature on matching markets and applications of matching theory to cognitive radio networks.  {Section \ref{sec:dry}  introduces} the context-free matching problem, the proposed BLMA and proof of convergence.  {Section \ref{sec:cog}  considers an application of the BLMA to cognitive radio networks assuming quasi-convex utilities.  {Section \ref{sec:modify} presents another application} for the BLMA for general nonlinear utilities.  {Section \ref{sec:num} presents} some numerical examples to {illustrate} our results and to compare with existing work in the literature. {Finally,} Section \ref{sec:conc} concludes this paper.% The symbols used in this paper are listed in Table \ref{symbol}.

%%%%%%%%%%%%%%%%%%%%%%%%%%%%%%%%%%%%%%%%%%%%%%%%%%%%%%%%%%%%%%
\section{Related Literature}
\label{sec:lit}
%%%%%%%%%%%%%%%%%%%%%%%%%%%%%%%%%%%%%%%%%%%%%%%%%%%%%%%%%%%%%%
\subsection{Related Work in Matching Theory}
%%%%%%%%%%%%%%%%%%%%%%%%%%%%%%%%%%%%%%%%%%%%%%%%%
Starting with two models, the marriage problem \cite{gale} and the assignment game \cite{shsh}, the study of stable matchings has evolved into a solid theory with applications in many areas \cite{dutting}. The effective coalitions in these games are two-player coalitions\cite{serr2}. The outcome of the assignment game is a specification of utilities to agents together with a matching such that every agent receives his most preferred match at the announced utilities \cite{alaei}. In the linear model in \cite{shsh}, generally, only one optimal matching exists that maximizes the sum of the utilities, and it is compatible with infinitely-many stable payoffs. In the model with general nonlinear utilities, first described by Demange and Gale in \cite{dg}, social welfare (i.e. maximizing the sum of the utilities) is not {well defined} since different agents' utilities are not measured in the same units. In this case, possibly several optimal matchings exist \cite{roth_book}. 

Numerous techniques exist on finding equilibrium, i.e. a matching supported by a stable payoff, in linear assignment games under various information assumptions \cite{np, chen,klaus}. Of particular interest to us is the work of Nax and Pradelski in \cite{np}. {This work} proposed a decentralized algorithm to find a pairwise stable solution in the linear assignment game involving matching workers and firms. The authors considered a limited information scenario so that market participants know nothing about other players' utilities, nor can they deduce such information from prior rounds of play. Agents have aspiration levels that they adjust from time to time based on their experienced payoffs. The algorithm proceeds by making random encounters between agents on both {sides} of the market. Agents then make offers to each other compatible with their current aspirations. If they both find their offers profitable, they {match;} otherwise they return to their old match or lower their aspirations if they were single. This dynamic learning algorithm is shown to converge to pairwise stable solutions in finite time with probability one. 

Building on this idea, we consider a similar dynamic for our matching problem which encompasses {\emph{both}} the linear and non-linear utility cases and show that this dynamic converges to stable solutions with probability one. In addition to the generalized nature of our matching formulation, our problem is enriched by the presence of many degrees of freedom that control the value of the utility\footnote{There are as many degrees of freedom as there are resources.}. This feature also enlarges the search space for pairwise stable solutions. We show however that such issues can be significantly suppressed by carefully designing the agents' negotiation mechanisms.

\subsection{Application of Matching theory in Cognitive Radio Networks}
%%%%%%%%%%%%%%%%%%%%%%%%%%%%%%%%%%%%%%%%%%%%%%%%%%%%%%%%%%%%%%%%%%%%%%%

In addition to the recent publication of \cite{mpusu1}, a growing body of literature is using matching theory to solve resource allocation problems in wireless communications, e.g. see \cite{mpusu1} for a comprehensive literature review, \cite{saad1} for a tutorial,  and \cite{fengdistmat, matching2,matching3, wiassign} and the references therein for various applications of matching theory to wireless networks. 

The closest work to ours is \cite{fengdistmat} where the scenario of multiple PUs and SUs is considered. SUs use part of their power to relay the {PU messages} in exchange for spectrum access time. The problem formulation permits the classification of SUs based on their so called type information, which is a compact representation of an SU's private information such as transmitter power, sensitivity to power {consumption,} and SU's transmitter to PU's receiver channel gain. The {paper proposes} a distributed algorithm to solve the matching problem under two assumptions on the available information. In the partially incomplete information scenario, PUs have knowledge of the types of all SUs connected to them, while in the incomplete information scenario, a PU knows the set of SUs' types connected to itself, but does not know the exact type of each SU. Note that knowledge of type information permits a PU to know the utility of an SU for a given time and power allocation. The {paper then compares} the performance of the two distributed algorithms in terms of PUs' and SUs' utilities. 

We consider a similar setting to \cite{fengdistmat} where SUs relay {PU} data in exchange for spectrum access time. However, we adopt less restrictive information assumptions. In our formulation, agents, whether PUs or SUs, have enough information to calculate their own utilities. No knowledge is available about the utilities of other users, whether on the same side or the opposite side of the market. The functional form of the utility of the SUs in our application does not permit type classification and we will show that such information will not be needed to reach desired solutions.

%%%%%%%%%%%%%%%%%%%%%%%%%%%%%%%%%%%%%%%%%%%%%%%%%
\section{Context-free BLMA Algorithm}
\label{sec:dry}

\subsection{Setup}

We consider a two-sided matching problem {constructed} as follows. There are two disjoint sets of agents,  ${\cal K}=\{1,2,...,K\}$ and ${\cal L}=\{1,2,...,L\}$, that form two sides of a matching market. We exclusively will use $k$ and $\ell$ to denote a representative element of $\mathcal{K}$ and $\mathcal{L}$, respectively, and sometimes use $j$ to denote a representative element of $\mathcal{K}\cup \mathcal{L}$.

\begin{definition} \textit{A \textbf{matching} is a mapping 
$$\mu : \mathcal{K}\cup \mathcal{L} \rightarrow \mathcal{K}\cup \mathcal{L}\cup \{\emptyset\}$$
such that for any $k\in\mathcal{K}$ and $\ell\in \mathcal{L}$:
\begin{itemize}
\item $\mu(k)\in \mathcal{L} \cup  \{\emptyset\}$.
\item $\mu(\ell)\in \mathcal{K}\cup  \{\emptyset\}$.
\item $\ell = \mu(k) \Leftrightarrow k = \mu(\ell)$.
\end{itemize}}
\end{definition}
If $\ell = \mu(k)$, then $k$ and $\ell$ are said to be \textit{matched}. In our model, an agent can be matched to at most one agent on the opposite side of the market. If $\mu(j) = \emptyset$, then agent $j\in \mathcal{K}\cup \mathcal{L}$ is said to be \textit{single}. 

A matching, $\mu$, can be characterized by a $K\times L$ matrix, $\mathbf{M}_\mu$, with elements in $\{0,1\}$, such that
$$\mathbf{M}_\mu(k,\ell) = 
\begin{cases}
1,&\mu(k) = \ell;\\
0,&\text{otherwise.}
\end{cases}$$
Let $\mathcal{M}$ denote the set of all feasible matching matrices induced by some matching, $\mu$. For any $\mathbf{M}\in\mathcal{M}$, let $\mu_\mathbf{M}$ denote the matching consistent with $\mathbf{M}$.

In order to describe the preferences of agents, we introduce \textit{aspiration levels} that abstractly represent the potential utility to be derived from a match. If $a_k$ is the aspiration level of agent $k\in\mathcal{K}$ and $b_\ell$ is the aspiration level of agent $\ell\in\mathcal{L}$, then
agents $k$ and $\ell$ are willing to be matched if the matching can produce utilities of at least $a_k$ and $b_\ell$, respectively.

More formally, we introduce the notion of an agreement function as follows.

\begin{definition} \textit{An \textbf{agreement function} is a mapping
$$\mathcal{A}:\mathbb{R}_+\times \mathbb{R}_+\rightarrow \{0,1\}$$
such that
\begin{enumerate}
\item If $\mathcal{A}(a,b) = 0$, then $\mathcal{A}(a',b')=0$ for all $a'\ge a$ and $b'\ge b$.
\item There exists a $\gamma > 0$ such that $\mathcal{A}(a,b) = 0$ if $a\ge \gamma$ or $b \ge \gamma$.
\end{enumerate}}
\end{definition}
We associate $\mathcal{A}(a,b) = 1$ to mean that the aspiration levels $a$ and $b$ are agreeable. Accordingly, condition 1 defines a monotonicity property for aspiration levels: once aspirations are not agreeable, further increases in aspiration levels also are not agreeable. Condition 2 defines a boundedness property for agreeable aspiration levels.

\begin{definition} \textit{A \textbf{matching problem} is a collection of agreement functions, $\mathcal{A}_{k\ell}$, indexed by $k\in \mathcal{K}$ and $\ell \in \mathcal{L}$.}
\end{definition}

We are interested in defining the notion of a stable outcome of a matching problem specified  by a set of agreement functions. Towards this end, we will consider vectors of aspiration levels $\mathbf{a}\in \mathbb{R}_+^K$ and $\mathbf{b}\in\mathbb{R}_+^L$, with elements denoted by $a_k$ and $b_\ell$, respectively. In the same way that $j$ denotes a representative element of $\mathcal{K}\cup \mathcal{L}$, we will use $c_j$ to denote the associated aspiration level. 

The specific stability notion of interest here will be $\epsilon$-pairwise stability defined as follows.

\begin{definition}\textit{
For $\epsilon > 0$, the matching $\mu$ and aspiration levels $\mathbf{a}$ and $\mathbf{b}$ form an $\epsilon$-\textbf{pairwise stable} solution to a matching problem if:
\begin{enumerate}
\item For all $(k,\ell)$ such that $\ell = \mu(k)$,
$$\mathcal{A}_{k\ell}(a_k,b_\ell) = 1.$$
\item For all $(k,\ell)$,
$$\mathcal{A}_{k\ell}(a_k+\epsilon,b_\ell+\epsilon) = 0.$$
\item For all $j\in \mathcal{K}\cup \mathcal{L}$ with $\mu(j) = \emptyset$, $c_j = 0$.
\end{enumerate}}
\end{definition}
In words, condition 1 states that aspiration levels between matched pairs are agreeable. Condition 2 implies that no pair of  agents have agreeable $\epsilon$-improvement aspiration levels. Note that condition 2 also applies to agents that are matched (i.e., even if $\ell = \mu(k)$). Condition 3 implies that single agents must {have zero aspiration levels at an $\epsilon$-pairwise stable solution. }

Here are two examples of matching problems using the above formulation. 
\begin{itemize}
\item \textit{Matching market with transferable utility:} The sets $\mathcal{K}$ and $\mathcal{L}$ represent firms and workers. For each pair, $k\in \mathcal{K}$ and $\ell\in \mathcal{L}$, 
the value $p_{k\ell}$ is the maximum salary firm $k$ is willing to pay worker $\ell$. Similarly, $q_{k\ell}$ is the minimum salary 
worker $\ell$ is willing to take to work for firm $k$. Suppose firm $k$ has aspiration level $a_k$ and worker $\ell$ has aspiration level $b_\ell$.
Then a match is agreeable if $\mathcal{A}_{k\ell}(a_k,b_\ell) = 1$, where
$$\mathcal{A}_{k\ell}(a_k,b_\ell) = \begin{cases}
1,& p_{k\ell} - a_k \ge q_{k\ell} + b_\ell;\\
0,&\text{otherwise}.\end{cases}$$
The agreement function $\mathcal{A}_{k\ell}$ is fully characterized by parameters $p_{k\ell}$ and $q_{k\ell}$.
\item \textit{Matching market with non-transferable utility:} There are two commodities, $G$ and $H$. An agent $k\in \mathcal{K}$ has an initial endowment of $g_k > 0$ of commodity $G$, whereas an agent $\ell\in \mathcal{L}$ has an initial endowment of $h_\ell > 0$ of commodity $H$. Every agent $k\in\mathcal{K}$ has an indexed collection of utility functions,
$$u_{k\ell}(g,h): \mathbb{R}_+ \times \mathbb{R}_+ \rightarrow \mathbb{R}$$
that expresses how much it values $g$ of its own commodity $G$ with $h$ of commodity $H$ from agent $\ell$. Likewise, every agent $\ell\in\mathcal{L}$ has an indexed collection of utility functions,
$$v_{k\ell}(g,h): \mathbb{R}_+ \times \mathbb{R}_+ \rightarrow \mathbb{R}$$
that expresses how much it values $h$ of its own commodity $H$ with $g$ of commodity $G$ from agent $k$. We assume that all utility functions are strictly increasing in both arguments. Suppose agent $k$ has aspiration level $a_k$ and {agent} $\ell$ has aspiration level $b_\ell$. Define the set
\begin{equation}
S_{k\ell}(a_k,b_\ell) = \theset{(g,h) \st g \le g_k; h \le h_\ell; u_{k\ell}(g_k - g,h) \ge a_k; v_{k\ell}(g,h_\ell - h) \ge b_\ell}.
\label{set}
\end{equation}
\noindent In words, this set describes all possible exchanges of $g$ from agent $k$ to agent $\ell$ in return for $h$ from agent $\ell$ to agent $k$ such that their utilities meet the specified aspiration levels. Then a match is agreeable if $\mathcal{A}_{k\ell}(a_k,b_\ell) = 1$, where
\begin{equation}
\mathcal{A}(a_k,b_\ell) = \begin{cases}
1,& S_{k\ell}(a_k,b_\ell) \not= {\{}\emptyset{\}};\\
0,&\text{otherwise.}\end{cases}
\label{agree}
\end{equation}
\end{itemize}

\subsection{BLMA Algorithm}

We now present an algorithm that leads to an $\epsilon$-pairwise stable solution. The algorithm is inspired by the recent work of \cite{np} on transferable utility assignment games.

Informally, the algorithm proceeds as follows:
\begin{itemize}
\item Aspirations levels, $\mathbf{a}(t)$ and $\mathbf{b}(t)$, as well as a matchings characterized by a matching matrix, $\mathbf{M}(t)$, evolve over stages $t=0,1,2,...$.
\item At stage $t$, a pair of agents, $(k,\ell)$, are activated at random.
\item If the increased aspiration levels $a_k(t)+\epsilon$ and $b_\ell(t)+\epsilon$ are agreeable, i.e., 
$$\mathcal{A}_{k\ell}(a_k(t)+\epsilon,b_\ell(t)+\epsilon) = 1,$$
then agents $k$ and $\ell$ become matched with { a positive probability $\eta$} and break previous matches, if any. The new aspiration levels $a_k(t+1)$ and $b_k(t+1)$, as well as matching matrix, $\mathbf{M}(t+1)$, are updated accordingly.
\item If the increased aspiration levels $a_k(t)+\epsilon$ and $b_\ell(t)+\epsilon$ are not agreeable, i.e., 
$$\mathcal{A}_{k\ell}(a_k(t)+\epsilon,b_\ell(t)+\epsilon) = 0,$$
then 
\begin{itemize}
\item The matching matrix remains unchanged, i.e., $\mathbf{M}(t+1) = \mathbf{M}(t)$.
\item If either agent $k$ or $\ell$ is single, then that agent reduces its aspiration by $\delta$, and the 
new aspiration levels $a_k(t+1)$ and/or $b_\ell(t+1)$ are updated accordingly.
\end{itemize}
\end{itemize}

Algorithm~\ref{alg:main} presents pseudocode for the Blind Matching Algorithm (BLMA). Here, the aforementioned ``stages'' are executions of the main loop. The time indexing of $t=0,1,2,...$ is suppressed for clarity of presentation. The notation ``$\textsc{rand}[0,1]$'' (Line 5) means an i.i.d.\ sample of a uniformly distributed random variable over the interval $[0,1]$. Also $[x]^+=\max(0,x)$ (Line 16).

BLMA is ``blind'' in the sense that potential matches between agents $k$ and $\ell$ are outcomes of bilateral negotiations that only depend on the agreement function $\mathcal{A}_{k\ell}$. The negotiation process is abstracted through the randomized outcome determined by $\textsc{rand}[0,1] \ge \eta$. An agent need not know the details behind another agent's acceptance or rejection. Furthermore, since such outcomes can be randomized, it may be difficult to make deterministic conclusions from a rejected offer. The main point is that all such issues are suppressed, with the specifics to depend on the actual context. Also, BLMA is non-deterministic in that revised aspiration levels (Lines 9--11) are not fully specified. Again, how this selection actually occurs will depend on the specific context.

\begin{algorithm}
\caption {BLMA}
\label{alg:main}
\begin{algorithmic}[1]
\REQUIRE $\epsilon > \delta > 0$ and $\eta \in (0,1]$.
\STATE Initialize $\mathbf{a} \ge 0$, $\mathbf{b}\ge 0$, $\mathbf{M} = \mathbf{0}$.
\LOOP
\STATE {Activate a pair of agents uniformly at random, $(k,\ell)\in\mathcal{K}\times\mathcal{L}$.}
\IF {${\cal A}_{kl}({ a_k+\epsilon},{ b_\ell+\epsilon})=1$}
\IF {$\textsc{rand}[0,1] \ge \eta$}
\STATE {${\mathbf M}(k,\ell)\leftarrow 1$}
\STATE {$\mathbf{M}(k,\ell')\leftarrow 0, \forall \ell'\not= \ell$}
\STATE {$\mathbf{M}(k',\ell)\leftarrow 0, \forall k'\not= k$}
\STATE {Select arbitrary $a'\ge a_k+\epsilon$ and $b'\ge b_\ell + \epsilon$ such that $\mathcal{A}_{k\ell}(a',b')=1$.}
\STATE {$a_k \leftarrow a'$}
\STATE {$b_\ell \leftarrow b'$}
\ENDIF
\ELSE
\FOR {$j\in \theset{k,\ell}$}
\IF {$\mu_\mathbf{M}(j) = \emptyset$,}
\STATE {$c_j \leftarrow [c_j - \delta]^+$}
\ENDIF
\ENDFOR
\ENDIF
\ENDLOOP
\end{algorithmic}
\end{algorithm}

We now state the main result.

\begin{theorem}\label{thm:main}\textit{
From any initial $\mathbf{a}\ge 0$, $\mathbf{b}\ge 0$, and $\mathbf{M} = 0$, the matching, $\mu_\mathbf{M}$, and aspiration levels, $\mathbf{a}$ and $\mathbf{b}$, converge to an $\epsilon$-pairwise stable matching with probability one.}\end{theorem}

\subsection{Proof}

This subsection is devoted to the proof of Theorem~\ref{thm:main}. We first introduce some specialized notation and terminology. 

We will use $z$ to denote the state of the algorithm. A state is a triplet
$$z \in \mathbb{R}_+^K\times \mathbb{R}_+^L \times \mathcal{M},$$
which is a combination of aspiration levels and a matching matrix. Each execution of the main loop results in an update in the state, e.g., $z \leftarrow z_\text{new}$. A state is \textit{reachable} if it can be realized in a finite number of main loop executions.

A state, $z= (\mathbf{a},\mathbf{b},\mathbf{M})$, will be called \textbf{pre-stable} if 
\begin{enumerate}
\item For all $(k,\ell)$ such that $\ell = \mu(k)$,
$$\mathcal{A}_{k\ell}(a_k,b_\ell) = 1.$$
\item For all $(k,\ell)$,
$$\mathcal{A}_{k\ell}(a_k+\epsilon,b_\ell+\epsilon) = 0.$$
\end{enumerate}
Strictly speaking, this definition depends on $\epsilon$ and accordingly could be called $\epsilon$-pre-stable. We will suppress this dependence for clarity of presentation. 

Note that the conditions for a pre-stable state are the first two conditions for an $\epsilon$-pairwise stable state. The only distinction is that a pre-stable state may have  single agents with non-zero aspiration levels.

\begin{claim}\label{claim:prestable}\textit{
From any reachable state, $z$, there exists a finite sequence of admissible transitions to a state, $z'$, that is pre-stable.
}\end{claim}
\proof\  In the main loop, activate and match any pair of agents $(k,\ell)$ with 
$${\cal A}_{kl}({ a_k+\epsilon},{ b_\ell+\epsilon})=1.$$
Continue to do so until there are no remaining such pairs. This process must terminate because of the boundedness property  of agreement functions and the algorithmic property that, in such a sequence, no agents are reducing their aspiration levels. Upon termination, the resulting state, $z'$, must be pre-stable by construction.
\QED

Given any state $z = (\mathbf{a},\mathbf{b},\mathbf{M})$, let $\text{SNZ}(z)$ denote the set of all single agents with non-zero aspiration levels, i.e.,
$$\text{SNZ}(z) = \theset{ j \in \mathcal{K}\cup \mathcal{L} \st \text{(i) } \mu(j) = \emptyset, \text{(ii) } c_j > 0}.$$

A pre-stable state, $z$, is called \textbf{tight} for any agent $j^*\in \text{SNZ}(z)$ with aspiration level $c_{j^*}$, {if} the new state, $z' = (\mathbf{a}',\mathbf{b}',\mathbf{M}')$, defined by:
$$\mathbf{M}' = \mathbf{M}$$
$$c_j' = \begin{cases}
c_j,&j\not= j^*;\\
[c_j - \delta]^+, & j = j^*,\end{cases}$$
is \textit{not} pre-stable. The implication here is that a state, $z$, is pre-stable and tight if (i) there are no $\epsilon$-improvement agreeable matches at current aspirations levels, but (ii) there will exist an $\epsilon$-improvement agreeable match after any agent in $\text{SNZ}(z)$ lowers its aspiration level by $\delta$.

\begin{claim}\label{claim:tight}\textit{
From any pre-stable state, $z$, there exists a finite sequence of admissible transitions to a state, $z'$, that is either (i) pre-stable and tight  or (ii) $\epsilon$-pairwise stable.}
\end{claim}
\proof\  Let $z$ be a pre-stable state. If $\text{SNZ}(z)$ is empty, then $z$ is already $\epsilon$-pairwise stable. Otherwise, select an arbitrary $j^*\in \text{SNZ}(z)$. In the main loop, let $j^*$ be activated with a matched agent. Since $z$ is pre-stable, the proposed match is not $\epsilon$-improvement agreeable, and so the new match is not accepted. Accordingly, the aspiration level $c_{j^*}$ is reduced by $\delta$. Repeat this sequence with the same agent $j^*$ until either $c_{j^*}$ is within a single $\delta$ reduction of admitting an $\epsilon$-improvement match (with some unspecified agent) or $c_{j^*} = 0$. Let $z^+$ be the resulting state. The only difference between $z$ and $z^+$ is in the aspiration level of $j^*$. Now select a different $j^{**}\in \text{SNZ}(z)$ and repeat accordingly. Upon visiting all of the agents in $\text{SNZ}(z)$, the resulting state $z'$ is pre-stable and tight by construction.
\QED

\begin{claim}\label{claim:final}\textit{From any pre-stable and tight state, $z$, with $\magn{\text{SNZ}(z)}\not= 0$, there exists a finite sequence of admissible transitions to a pre-stable state $z'$ with $\magn{\text{SNZ}(z') } < \magn{\text{SNZ}(z)}$, i.e., a strict reduction in the number of single agents with non-zero aspiration levels.}
\end{claim}
\proof\ Let $z$ be a state that is pre-stable and tight. Let us assume without loss of generality that there exists a $k^*\in \text{SNZ}(z)$, i.e., some agent in $\mathcal{K}$ is single with non-zero aspiration levels. (Analogous arguments hold if the selected agent is in $\mathcal{L}$.) Let us call $k^*$ the token holding agent. The token will not be released until
a new state, $z'$, is reached with the desired reduction in cardinality.

Activate (the token holding) $k^*$ with any matched agent $\ell\in\mathcal{L}$.  Since $z$ is pre-stable, a new match does not occur, and the state is updated so that agent $k^*$ has an aspiration level of $a_{k^*} - \delta$.\footnote{Here, we assume for convenience that $a_{k^*} - \delta > 0$. Similar arguments hold in case $[a_{k^*} - \delta]^+ = 0$.}
Furthermore, since $z$ was tight, there exists an agent $\ell^*$ for which
$$\mathcal{A}_{k^*\ell^*}(a_{k^*} - \delta+\epsilon,{b}_{\ell^*}+\epsilon) = 1.$$
Let the match between $k^*$ and $\ell^*$ occur. There are three possible scenarios:
\begin{itemize}
\item \textit{Scenario A.  $\ell^*\in \text{SNZ}(z)$:} Agent $\ell^*$ was also single with non-zero aspiration. Let $z^+$ denote the resulting state, and $a_{k^*}^+$ and $b_{\ell^*}^+$ be the revised aspiration levels of agents $k^*$ and $\ell^*$, respectively. Then
\begin{gather*}
a_{k^*}^+ \ge a_{k^*} - \delta + \epsilon,\\
b_{\ell^*}^+ \ge b_{\ell^*} + \epsilon.
\end{gather*}
Furthermore, since $\epsilon > \delta$, the aspiration levels of $z^+$ are greater than the aspiration levels of $z$. Accordingly, $z^+$ is pre-stable. Now apply the procedure of Claim~\ref{claim:tight} to produce a state $z'$ that is pre-stable and tight. By construction, the number of single agents with non-zero aspiration has been reduced by at least two, i.e.,
$${|}\text{SNZ}(z'){|}\le {|}\text{SNZ}(z){|}- 2.$$
Accordingly, the token is released.

\item \textit{Scenario B.  $\mu(\ell^*) = \emptyset$ and $b_{\ell^*}= 0$:} Agent $\ell^*$ is also single, but with zero aspiration level. Proceed in a similar manner to Scenario A to construct a state $z'$ where the number of single agents with non-zero aspiration has been reduced by at least one, i.e., 
$${|}\text{SNZ}(z'){|}\le {|}\text{SNZ}(z){|}- 1.$$
Accordingly, the token is released.

\item \textit{Scenario C. $\mu(\ell^*)  = k^{**}$:} Agent $\ell^*$ is matched to another agent, namely $k^{**}$, on the $\mathcal{K}$ side of the market. Increase the aspiration levels of newly matched agents $k^*$ and $\ell^*$ as required. Furthermore, reassign the token to the newly single agent, $k^{**}$. Note that at this stage, the number of single agents with non-zero aspiration levels has not changed, and hence the token has not been released, but rather reassigned. The new state would be pre-stable and tight except for the aspiration level of the new token holding agent, $k^{**}$. Accordingly, through a series of executions of the main loop, reduce the aspiration level of $k^{**}$ until either $c_{k^{**}} = 0$ or the realized state is pre-stable and tight. In the former case, the number of single agents with non-zero aspiration levels has been reduced by one, as desired, and the token is released. In the latter case, since the realized state is pre-stable and tight, one can now invoke the aforementioned procedures in the proof of Claim~3 \textit{while selecting $k^{**}$ as the token holding agent} (i.e., without reassigning the token). If the outcome is Scenario A or B, then the number of singe agents with non-zero aspirations levels has been reduced, and the token is released. Otherwise, the token is passed to yet another agent, e.g., $k^{***}$, and the process is repeated. Note that whenever the token is reassigned, it stays on the same side, $\mathcal{K}$, of the market. Furthermore, with each reassignment, the sum of the aspiration levels of the $\mathcal{L}$ side of the market strictly {increases}. Such {increase} cannot continue indefinitely {because of the boundedness of the agreement functions}, and so eventually the token must be released with the number of single non-zero agents reduced. Now apply the procedure of Claim~\ref{claim:tight} to assure that the exiting state $z'$ is pre-stable and tight. 
\end{itemize}
\QED

With Claims~\ref{claim:prestable}--\ref{claim:final} in place, we are now in a position to prove Theorem~\ref{thm:main}. From any reachable state, there exists a finite sequence of admissible transitions that leads to a pre-stable state (Claim~\ref{claim:prestable}) followed by a finite sequence of admissible transitions that leads to a pre-stable and tight state (Claim~\ref{claim:tight}). By a repeated application of Claim~3, there exists a finite sequence of admissible transitions to an $\epsilon$-pairwise stable state. 

\begin{figure}
\begin{center}
\includegraphics[width=0.4\textwidth]{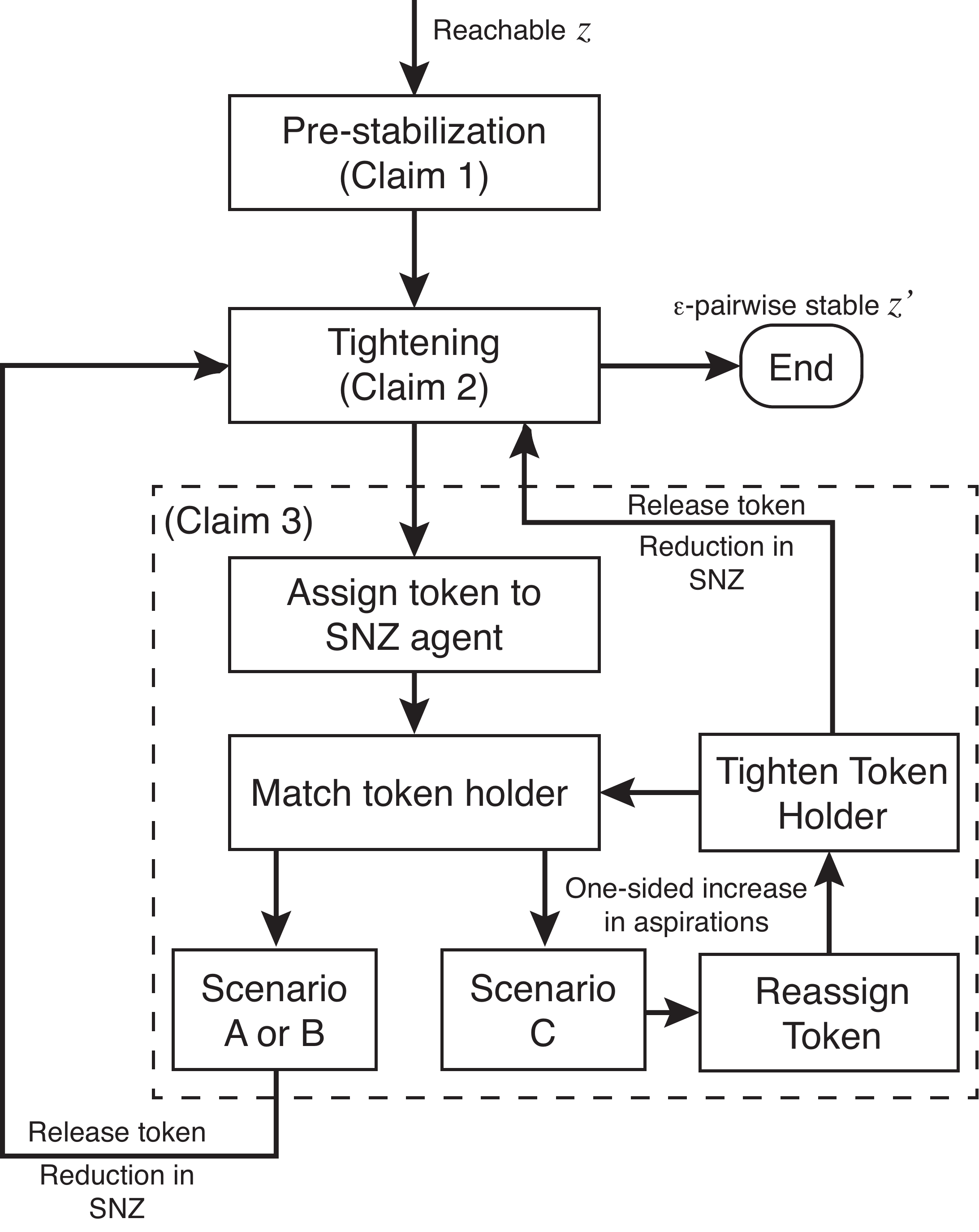}
\end{center}
\caption{Illustration of the combined effects of Claims~\ref{claim:prestable}--\ref{claim:final} to transition from any reachable state, $z$, to an $\epsilon$-pairwise stable state (``End'').}\label{fig:proof}
\end{figure}

Fig.~\ref{fig:proof} illustrates the combined effect of Claims~\ref{claim:prestable}--\ref{claim:final}. A state is first made to be pre-stable and then pre-stable and tight. At that point, the procedure behind Claim~\ref{claim:final} is executed. There are two types of loops in Figure~\ref{fig:proof}. The first type of loop involves a \textit{releasing} of the token and a return to the ``Tightening'' procedure. This loop results in a reduction in the number of single agents with non-zero aspiration levels, and so there can only be a finite number of such iterations. The second type of loop involves a \textit{reassignment} of the token. In this loop,  the number of single agents with non-zero aspiration levels remains constant. However, this loop results in an increase in the sum of aspiration levels on one side of the market (namely, on the side opposite to the token), and so this loop must eventually be exited because of the boundedness assumption on agreement functions. Ultimately, the process much reach the ``End'' state, which is $\epsilon$-pairwise stable. 

Note that Figure~1 is intended to illustrate a feasible sequence  of admissible transitions to an $\epsilon$-pairwise stable state, and as such a \textit{positive probability} flow of the BLMA algorithm. Putting all this together, the conclusion is that from any reachable state, there is a positive probability of a finite sequence of admissible transitions eventually leading to an $\epsilon$-pairwise stable state. Given the finite cardinality of $\mathcal{K}$ and $\mathcal{L}$, and the boundedness property of agreement functions, there exists some finite number of executions of the main loop, say $T$, such that from any reachable state, $z$, the probability of reaching an $\epsilon$-pairwise stable state after $T$ executions of the main loop is at least $p > 0$, where both $p$ and $T$ do not depend on $z$. By the Borel-Cantelli lemma, we can subsequently conclude that the algorithm reaches an $\epsilon$-pairwise stable state with probability one \cite{papoulis}.

%In the remainder of this paper, we will consider different negotiation mechanisms that we will show can be different realizations of the BLMA and a way of implementing it in various applications. 
%%%%%%%%%%%%%%%%%%%%%%%%%%%%%%%%%%%%%%%%%%%%%%%%%

%%%%%%%%%%%%%%%%%%%%%%%%%%%%%%%%%%%%%%%%%%%%%%%%%%%%%%%%%%%%%
\section{Application: Cognitive Radio Market with Quasi-convex Utilities}
%%%%%%%%%%%%%%%%%%%%%%%%%%%%%%%%%%%%%%%%%%%%%%%%%%%%%%%%%%%%%
\label{sec:cog}

In this section, we apply the BLMA to solve a matching problem in cognitive radio networks. We consider overlay spectrum access in cognitive networks whereby primary users (PUs) allow secondary users (SUs) access to their spectrum in exchange for some compensation, such as money or resource \cite{gold, zhao07}. Specifically, we will use cooperative spectrum sharing as one such dynamic spectrum access technique in which SUs relay traffic for PUs in exchange for dedicated spectrum access time for their own communication\cite{pumsu}. Previous work also focused on designing cooperative spectrum access techniques for PUs and SUs {(e.g. \cite{mpusu1, fengdistmat})}. We will adopt the system model of \cite{fengdistmat} in this section and provide a solution using the BLMA with significantly less information assumptions.  

{\subsection{System Model}}

Consider a cognitive radio network comprised of a set ${\cal K}=\{1,2,...,K\}$ {of} PUs and a set ${\cal L}=\{1,2,...,L\}$ {of} SUs. Each network node is made up of a transmitter-receiver pair. The PUs are the owners of the {network,} and they control how the spectrum is accessed. The SUs are opportunistic users that seek possible spectrum {access to the PU network with agreed-upon time and power allocations.} {Suppose that $\text{PU}_k$ is matched with $\text{SU}_\ell$ with agreed upon time, $\tau_{k\ell}\in [0,1]$, and power, $P_{k\ell} > 0$. The} $\text{PU}_k$'s time slot, $\tau_k$, is fixed for cooperative communication while $\tau_{kl}$ is the time allocated by $\text{PU}_{kl}$ to $\text{SU}_\ell$'s communications. The three phases of communication are as follows:
\begin{itemize}
\item During the first $\frac{\tau_k}{2}$ part of the time slot, the PU's transmitter, PT$_k$, broadcasts its data packet. The data is received by the PU's receiver, PR$_k$, and by SU's transmitter, ST$_\ell$, contingent on $h_{k\ell}\geq h_{k}$, where $h_{k\ell}$ is PT$_k$ to ST$_\ell$ channel gain, while $h_{k}$ is PT$_k$ to PR$_k$ direct link channel gain.
\item During the second $\frac{\tau_k}{2}$ part of the time slot, ST$_\ell$ decodes the data received in phase 1 and relays $\text{PU}_k$'s message to PR$_k$ using power $P_{k\ell}$. 
\item A third time phase, ${\tau_{k\ell}}$, is allocated by $\text{PU}_k$ for $\text{SU}_\ell$'s own communication. 
\end{itemize}

\begin{figure}
	\centering
		\includegraphics[scale=0.4]{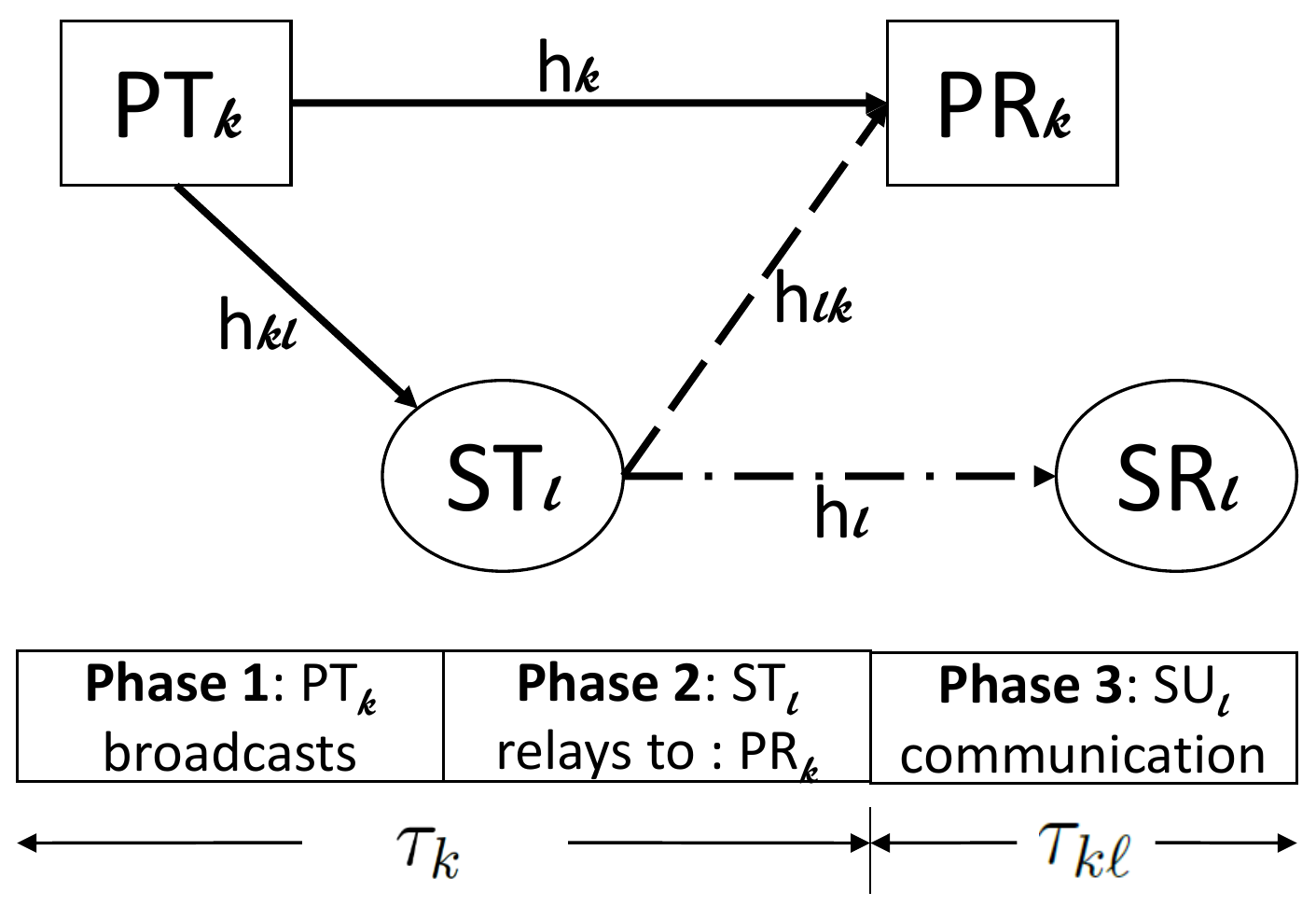}
	\caption{Interaction between $\text{PU}_k$ and $\text{SU}_\ell$ during the three transmission phases. }
	\label{fig:one}
\end{figure}

The above communication structure is shown in Fig. \ref{fig:one} with the other relevant channel gains. Once matched, $\text{PU}_k$ gives up time, $\tau_{k\ell}$, for $\text{SU}_\ell$'s spectrum access, in exchange for relaying help from $\text{SU}_\ell$. Moreover, $\text{SU}_\ell$ uses power, $P_{k\ell}$, for relaying $\text{PU}_k$'s message in exchange for spectrum access.

{\subsection{Utility Functions and Matching Problem Formulation}}

Suppose that specific time and power allocations have been agreed upon by $\text{PU}_k$ and $\text{SU}_\ell$. Assuming, without loss of generality, that $\tau_k=1$, then the average achievable $\text{PU}_k$ data rate over the entire time when matched with $\text{SU}_\ell$ is 

\begin{equation}
R^{\rm P}_{k\ell}(\tau_{k\ell},P_{k\ell})=\begin{cases}
\frac{1}{2\left(1+\tau_{k\ell}\right)}\left[R^{\rm P,dl}_{k}+\log_2\left(1+\frac{P_{k\ell}}{\sigma^2}\right)\right]&\mbox{  if  $h_{k\ell}\geq h_{k}$}\\
0&\mbox{  otherwise},
\end{cases}
\label{rp}
\end{equation}
\

\noindent where the above rate expression can be achieved using a decode-and-forward protocol with parallel channel coding \cite{lane,rate_add}. The requirement $h_{k\ell}\geq h_{k}$ is the condition that $\text{ST}_\ell$ is in the decoding set of $\text{PU}_k$\cite{decode}. $R^{\rm P,dl}_{k}$ is $\text{PU}_k$'s direct link rate which can be calculated as

\begin{equation}
R^{\rm P,dl}_{k}=\log_2\left(1+\frac{P_k|h_{k}|^2}{\sigma^2}\right),
\end{equation}

\noindent where $P_k$ is $\text{PU}_k$'s transmission power. Equation (\ref{rp}) {is comprised of} two terms. The first term, $R^{\rm P,dl}_{k}$, refers to phase one of the communication, while the second term reflects the rate accrued due to relaying by $\text{SU}_\ell$ in the second communication phase. Then, the associated utility for $\text{PU}_k$ when matched with $\text{SU}_\ell$ is
\begin{equation}
u_{k\ell}(\tau_{k\ell},P_{k\ell})=[R^{\rm P}_{k\ell}(\tau_{k\ell},P_{k\ell})-R^{\rm P,dl}_{k}]^+,
\label{up}
\end{equation}

\noindent where $u_{k\ell}\geq 0$ reflects the fact that a matching with any SU must at least provide the same rate as if $\text{PU}_k$ {were} single. We note in (\ref{rp}) that the matched SU offers the PU a fixed {rate;} i.e. $\text{SU}_\ell$ uses power $P_{k\ell}/|h_{\ell k}|^2$ to send $\text{PU}_k$'s packets so that the rate enhancement to $\text{PU}_k$ is kept constant at $\frac{1}{2}\log_2\left(1+\frac{P_{k\ell}}{\sigma^2}\right)$. 

{On the other side,} $\text{SU}_\ell$ gains utility from the spectrum access opportunity when matched with $\text{PU}_k$, while paying a relaying {\emph{price}}, in terms of power used to help the primary {link. The} secondary utility can be written as
\begin{equation}
v_{k\ell}(\tau_{k\ell},P_{k\ell})\!\!=\!\!\left[{\tau_{k\ell}}\log_2\left(\!\!1\!+\!\frac{{P}_{\ell}|h_{\ell}|^2}{\sigma^2}\!\!\right)-c_\ell\left(\frac{P_{k\ell}}{2|h_{lk}|^2}+\tau_{kl}P_\ell\right)\right]^+,
\label{us}
\end{equation}

\noindent where ${P}_{\ell}$ is the power used for $\text{SU}_\ell$'s own communication and $c_\ell$ is $\text{SU}_\ell$'s sensitivity for unit power consumption.  We are interested in finding a way to pair the PUs and the SUs with the appropriate time and power allocations so that no party has an incentive to break the match. We will cast the problem at hand in the matching problem formulation. 

%{Hereafter, to reduce notational clutter, we will omit the utility dependence on $P_{k\ell}$ and $\tau_{k\ell}$ where it is implied or clear.}

Recall the matching market with non-transferable utility introduced in the previous section. The two commodities here are time and power. Assume that any PU can allocate a maximum of $\tau_{k\ell}\leq 1$ for all $\ell\in{\cal L}$. Given equation (\ref{us}) and a specified $\tau_{k\ell}$ value, the SU can allocate a maximum power of $\overline{P}_{k\ell}(\tau_{k\ell})$. This is the $P_{k\ell}$ value which yields zero utility for $\text{SU}_\ell$ in (\ref{us}). Since we chose to write the utilities in terms of $\tau_{k\ell}$, i.e. the time allotted by $\text{PU}_{k\ell}$ to $SU_{k\ell}$ and $P_{k\ell}$, the power granted by $\text{SU}_\ell$ to $\text{PU}_k$, the set $S_{k\ell}(a_k,b_\ell)$ in (\ref{set}) can now be redefined as 
\begin{equation}
S_{k\ell}(a_k,b_\ell) = \theset{(\tau_{k\ell},P_{k\ell}) \st \tau_{k\ell} \leq 1; P_{k\ell} \leq \overline{P}_{k\ell}(\tau_{k\ell}); u_{k\ell}(\tau_{k\ell},P_{k\ell}) \ge a_k; v_{k\ell}(\tau_{k\ell},P_{k\ell}) \ge b_\ell}.
\label{set2}
\end{equation}

The agreement functions ${\cal A}_{k\ell}$ are defined as in (\ref{agree}) and the matching problem formulation follows accordingly. \\

%{The condition} in (\ref{stable2}) is sometimes referred to as the incentive compatibility (IC) constraint since satisfying it ensures that the matching will not be blocked by any PU-SU pair who can break off their assigned matches and receive a better payoff if paired together instead. There is another condition, called individual rationality (IR), which, if satisfied, guarantees that the payoff to any PU or SU will be large enough so that the given user prefers its allocated match rather than to stay single. Given the utility forms of (\ref{up}) and (\ref{us}), the utility of staying single is zero to both users. Thus the IR condition simplifies to $u_{k\ell}\geq 0$ and $v_{k\ell}\geq 0$ for the PU and the SU, respectively.% The condition in (\ref{stable2}) simply states that once two users are matched, the time and power resources are totally split between them.

%%%%%%%%%%%%%%%%%%%%%%%%%%%%%%%%%%%%%%%%%%%%%%%%%%%%%%%
{\subsection{ BLMA Realization}}

%%%%%%%%%%%%%%%%%%%%%%%%%%%%%%%%%%%%%%%%%%%%%%%%%%%%%%%%%%%%%%

Since the problem of assigning PUs to SUs with agreed upon time and power allocations is a matching problem in the sense defined in Section \ref{sec:dry}, we will be able to find an $\epsilon$-pairwise stable matching solution using the BLMA. We will illustrate here the details by which the BLMA will operate within the cognitive radio context. We note first from (\ref{up}) that $u_{k\ell}$ is decreasing in $\tau_{k\ell}$ and increasing in $P_{k\ell}$, while the opposite is true for $v_{k\ell}$. Furthermore, we can easily verify the utility $u_{k\ell}$ is quasi-convex and the utility $v_{k\ell}$ is quasi-concave\footnote{A function $f:\mathcal{K}\rightarrow \mathbb{R}$ over a convex set, $\mathcal{K}$, is quasi-concave (quasi-convex) if super-level (sub-level) sets , $\{x : f(x)\ge \rho\}$ ($\{x : f(x)\le \rho\}$), are convex \cite{boydconvex}.}\footnote{In fact $v_{k\ell}$ is only linear in $t_{k\ell}$ and $P_{k\ell}$ as can be verified from equation (\ref{us}) but we use the more general ``quasi-concave'' term to emphasize that the modified algorithm will still function under this less-strict assumption.}. Consider the following two procedures:

\begin{enumerate}\item\emph{BLMA1: The PU-SU negotiation process}

\begin{itemize}
\item Active agents make offers compatible with their aspirations levels.
\begin{enumerate} 
\item Let $\text{PU}_k$ pick, uniformly at random, an offer $(\tau^{\rm P}_{k\ell},P^{\rm P}_{k\ell})$ such that $a_k+\epsilon=u_{k\ell}(\tau^{\rm P}_{k\ell},P^{\rm P}_{k\ell})$.
\item Let $\text{SU}_\ell$ pick, uniformly at random, an offer $(\tau^{\rm S}_{k\ell},P^{\rm S}_{k\ell})$ such that $b_\ell+\epsilon=v_{k\ell}(\tau^{\rm S}_{k\ell},P^{\rm S}_{k\ell})$.
\end{enumerate}
\item \textbf{If}
\begin{enumerate} 
\item $\left\lfloor u_{k\ell}(\tau^{\rm S}_{k\ell},P^{\rm S}_{k\ell})\right\rfloor_\delta\geq a_k+\epsilon$, and
\item $\left\lfloor v_{k\ell}(\tau^{\rm P}_{k\ell},P^{\rm P}_{k\ell})\right\rfloor_\delta\geq b_\ell+\epsilon$
\end{enumerate}
\,\,\,\, \textbf{Then}\\
\,\,\,\,\,\,\, ${\cal A}_{k\ell}(a_k+\epsilon,b_\ell+\epsilon)=1$.\\
\textbf{End If}
\end{itemize}
\item \emph{BLMA1: Updating the aspiration levels}
\begin{itemize}
\item {Select point $(\tau_{k\ell}, P_{k\ell})$ uniformly at random on the line segment connecting $(\tau^{\rm P}_{k\ell}, P^{\rm P}_{k\ell})$ and $(\tau^{\rm S}_{k\ell}, P^{\rm S}_{k\ell})$.}
\begin{enumerate}
\item {Update $a_k \leftarrow \left\lfloor{ u^{\rm P}_{k\ell}(\tau_{k\ell}, P_{k\ell})}\right\rfloor_\delta$.}
\item {Update $b_\ell \leftarrow \left\lfloor{u^{\rm S}_{k\ell}(\tau_{k\ell}, P_{k\ell})}\right\rfloor_\delta$.}
\end{enumerate}
\end{itemize}
\end{enumerate}
\noindent To compare all the forthcoming realizations of the BLMA, we will collectively refer to the above two procedures as BLMA1. Note that $(\tau^{\rm P}_{k\ell},P^{\rm P}_{k\ell})$ is the time and power offer of $\text{PU}_k$, $(\tau^{\rm S}_{k\ell},P^{\rm S}_{k\ell})$ is the time and power offer of $\text{SU}_\ell$, and $\left\lfloor x\right\rfloor_\delta=\max\theset{m\delta\st m\delta\leq x \text{ for }m \in{\mathbb Z}_+}$. Considering the above two procedures, we have the following result: % Intuitively, ascertaining whether an agreement of aspiration levels exists proceeds as follows. Active agents randomly pick points on their aspiration curves, i.e. time and power pairs that yield the same aspiration level. These are the agents' offers. Each agent then calculates the minimum ``potential'' utility calculated at the received offer. If this potential utility is greater than the $\epsilon+$aspiration level for both agents, then there is mutual agreement and the match can be made.
 
\begin{proposition}
Given quasi-convex utility $u_{k\ell}$ and quasi-concave utility $v_{k\ell}$ and $\epsilon=q\delta$ for some $\theset{q \st q>1 \text{ and } q\in {\mathbb Z}_+}$, BLMA1 will converge to an $\epsilon$-pairwise stable state with probability one. 
\end{proposition}

\proof
Read the negotiation process above as Line 4 in the BLMA. $\text{PU}_k$ with aspiration level $a_k$ and $\text{SU}_\ell$ with aspiration level $b_\ell$ will only declare their match agreeable when $\left\lfloor u_{k\ell}(\tau^{\rm S}_{k\ell},P^{\rm S}_{k\ell})\right\rfloor_\delta\geq a_k+\epsilon=u_{k\ell}(\tau^{\rm P}_{k\ell},P^{\rm P}_{k\ell})$, and $\left\lfloor v_{k\ell}(\tau^{\rm P}_{k\ell},P^{\rm P}_{k\ell})\right\rfloor_\delta\geq b_\ell+\epsilon=v_{k\ell}(\tau^{\rm S}_{k\ell},P^{\rm S}_{k\ell})$. Read the aspiration update above as Line 9 in the BLMA. Quasi-convexity of the utilities ensures that any point on the line segment connecting $(\tau^{\rm P}_{k\ell}, P^{\rm P}_{k\ell})$ and $(\tau^{\rm S}_{k\ell}, P^{\rm S}_{k\ell})$ is agreeable. The revised aspiration levels will be $a_k^+\geq a_k+\epsilon>a_k$ and $b_\ell^+\geq b_\ell+\epsilon>b_\ell$ since $\epsilon>0$. The process cannot continue indefinitely by the boundedness of the aspiration levels and we will reach a pre-stable state, hence Claim 1 is established. {Fig. \ref{fig:2D1}} provides an illustration\footnote{Although $v_{k\ell}$ is linear in its variables, we plot $\text{SU}_\ell$'s aspiration contour as being quasi-concave function in Fig. \ref{fig:2D1} to assert that BLMA1 will still work for quasi-concave functions also.}. 

Claim 2 is established by Lines 15 and 16 and hence is not changed. Considering a pre-stable and tight state $z$, let $\text{SU}_\ell$ be any single in $\text{SNZ}(z)$ with aspiration $b_\ell$. Activate $\text{SU}_\ell$ with any agent $\text{PU}_k$ with aspiration $a_k$. There is no agreement since the state is pre-stable and tight. Since $\text{SU}_\ell$ is single, it lowers its aspiration level by $\delta$. Since the aspirations were tight, we now have an agreement. The two agents will match with revised aspirations $a_k^+\geq a_k+\epsilon>a_k$ and $b_\ell^+\geq b_\ell-\delta+\epsilon\geq b_\ell$ since $\epsilon>\delta>0$. Since the aspiration levels were pre-stable and tight and agents $\text{PU}_k$ and $\text{SU}_\ell$ increased their aspiration levels, we are again at a pre-stable state. We now consider the recursive application of Claim 3 coupled with the tightening process of Claim 2 to reach an $\epsilon$-pairwise stable state.

Finally, we make a note that $a_k,\,\,b_\ell\in\theset{ 0,\delta,2\delta,...,\gamma}$ by the requirement $\epsilon=q\delta$ and the ``flooring'' procedure of the aspiration update (line 9 in the BLMA). This confinement of the aspiration levels to the discrete grid of $\delta$ steps and the requirement of $\epsilon$-improvement agreements ensure that as the algorithm progresses and the aspiration levels pre-stabilize with agents squeeezing out all available resources that there still exists some $\eta>0$ probability of making the match (Line 5 in the BLMA).
 
\QED

\begin{figure}
	\centering
		\includegraphics[scale=0.55]{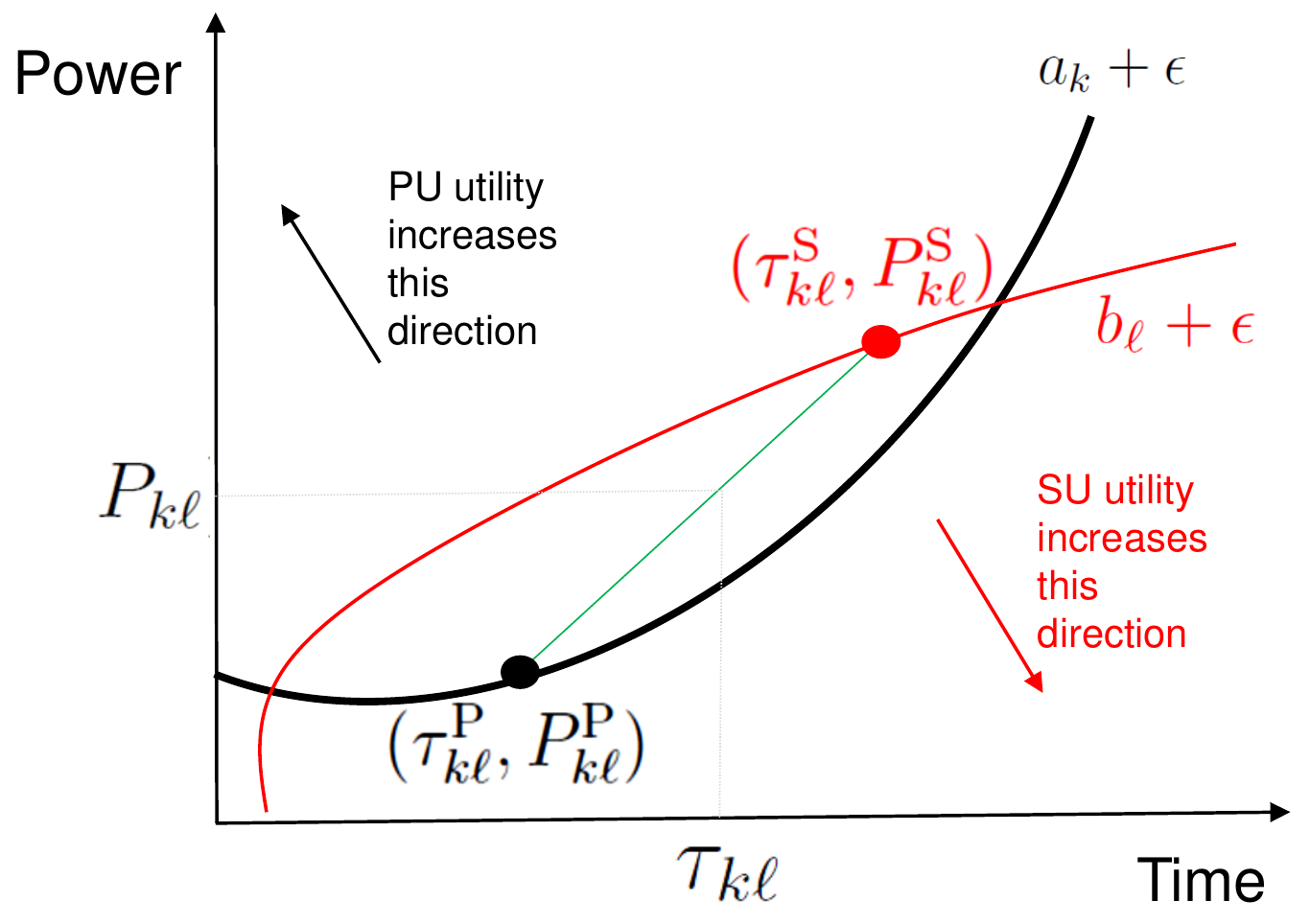}
	\caption{Figure shows contours for agents' aspiration levels given equations (\ref{up}) and (\ref{us}).  An agreement of aspiration levels $a_k+\epsilon$ and $b_\ell+\epsilon$ implies that the $b_\ell+\epsilon$ curve must {have some segment} above the $a_k+\epsilon$ curve, otherwise $u_{kl}(\tau^{\rm S}_{k\ell},P^{\rm S}_{k\ell}) <a_k+\epsilon$ or $v_{kl}(\tau^{\rm P}_{k\ell},P^{\rm P}_{k\ell}) <b_\ell+\epsilon$.  $\text{PU}_k$ will choose a random point on this $a_{k}+\epsilon$ contour, $( \tau^{\rm P}_{k\ell}, P^{\rm P}_{k\ell})$, as its offer to $\text{SU}_\ell$. $\text{SU}_\ell$ will also choose a random point on its $b_{\ell}+\epsilon$ contour, $(\tau^{\rm S}_{k\ell}, P^{\rm S}_{k\ell})$, as its offer to $\text{PU}_k$. The match point $(\tau_{k\ell}, P_{k\ell})$ will be chosen at uniformly at random on the line connecting these two offers.}
	\label{fig:2D1}
\end{figure}

\subsection{ Limited Information Scenario}

In this subsection, we motivate our choices for the PU-SU negotiation process and aspiration level update. Besides quasi-convexity, a careful inspection of the above procedures will reveal that we have only assumed that agents have all the information needed to calculate their own utility but no information is available about the utilities of other users on either side of the market. Fig. \ref{fig:info} illustrates the information users have about each other under the assumption of quasi-convex utility functions. Given a time and power offer from $\text{SU}_\ell$, $\text{PU}_k$ only has information of its aspiration contour $a_k$ and $\text{SU}_\ell$'s offer $(\tau^{\rm S}_{k\ell}, P^{\rm S}_{k\ell})$. $\text{PU}_k$ has no knowledge of any other points on the $b_{\ell}$ contour. Points on the line connecting the two offers $(\tau^{\rm P}_{k\ell}, P^{\rm P}_{k\ell})$ and $(\tau^{\rm S}_{k\ell}, P^{\rm S}_{k\ell})$ are agreeable from $\text{PU}_k$'s perspective given the quasi-convexity. A similar statement can be made about $\text{SU}_\ell$.

\begin{figure}
	\centering
		\includegraphics[scale=0.5]{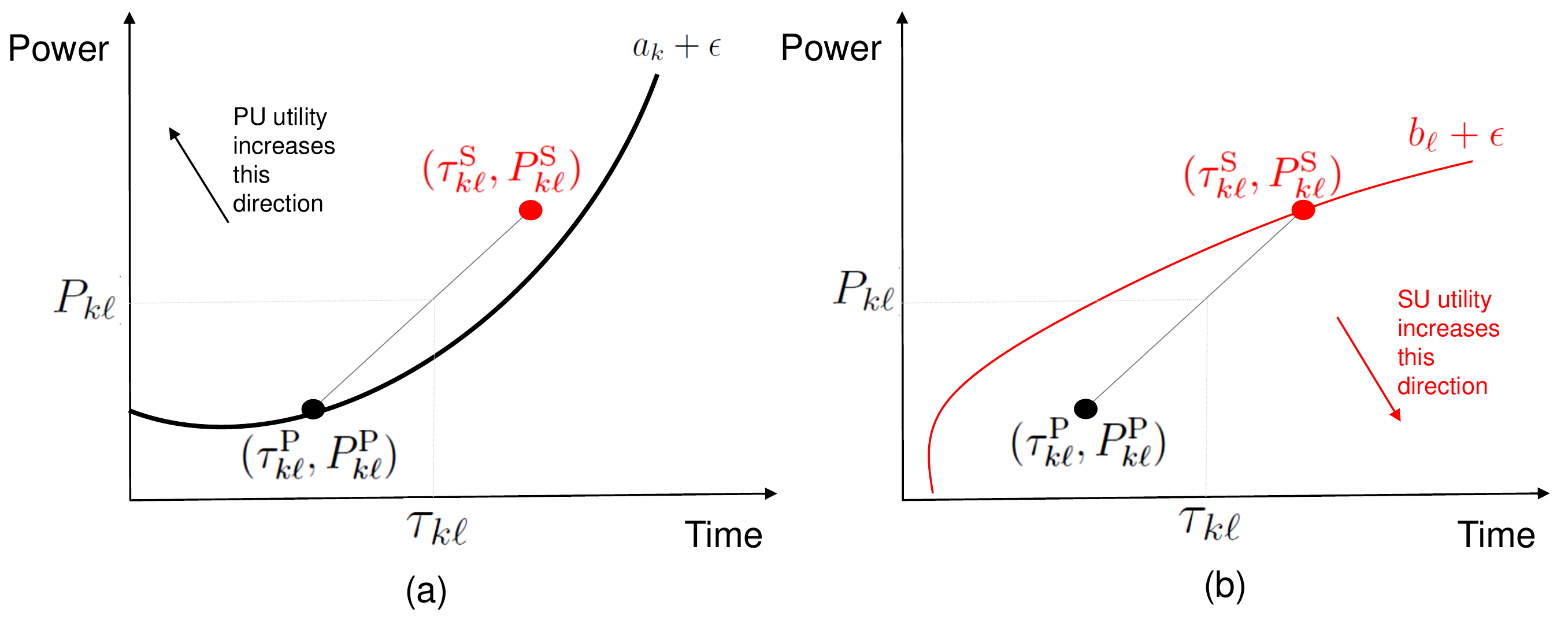}
	\caption{(a) Information available at $\text{PU}_k$ given $\text{SU}_\ell$'s offer and (b) Information available at $\text{SU}_\ell$ given $\text{PU}_k$'s offer. Given the functional trends for $\text{PU}_k$ and $\text{SU}_\ell$ with time and power, the straight line connecting the two offers guarantees an increase in utility for both users. }
	\label{fig:info}
\end{figure}

We contrast {this framework}, for example, with the work in \cite{fengdistmat}, wherein equation (\ref{us}) is normalized and its terms re-arranged so that the $\text{SU}_\ell$'s utility is of the form 
\begin{equation}
v^{\rm alt}_{k\ell}=\tau_{k\ell}\theta_{k\ell}-P_{k\ell},
\label{us2}
\end{equation}
\noindent and $\theta_{k\ell}$ is a compact representation of $\text{SU}_\ell$'s private information, or so called type, when paired with $\text{PU}_k$. The SU's type contains information as: 1) The relaying capability of $\text{SU}_\ell$ with $\text{PU}_k$, 2) the channel coefficients $h_{\ell k}$ and $h_\ell$, and 3) $\text{SU}_\ell$'s sensitivity for unit power consumption. The work in \cite{fengdistmat} considers two types of limited information scenarios. Under the partially incomplete information scenario, a PU knows the type of each SU connected to itself, but not to other PUs. Under the incomplete information scenario, a PU knows the values of {all the SU} types but has no way of associating a particular type with a specific SU\footnote{The authors in \cite{fengdistmat} do not motivate how such an incomplete information scenario can be realized.}. Given the above discussion, it is clear that we consider a more restrictive incomplete information scenario. It is also in that sense, that we name our algorithm a blind matching algorithm since very little information is available about other users in the market, yet we still converge to a stable outcome. We will contrast our approach with the one in \cite{fengdistmat} in the numerical results section.

%%%%%%%%%%%%%%%%%%%%%%%%%%%%%%%%%%%%%%%%%%%%%%%%%%%%%%%%%%%%%%
\section{BLMA Application to a Cognitive Radio Network with General Utilities}
\label{sec:modify}
%%%%%%%%%%%%%%%%%%%%%%%%%%%%%%%%%%%%%%%%%%%%%%%%%%%%%%%%%%%%%%

In this section, we will modify the system model slightly to showcase an application of the BLMA wherein the utilities of agents on the two sides of the market show opposite trends but are not necessarily quasi-convex\footnote{The realization suggested in this section naturally include the more stringent case of quasi-convex utilities.}. We will continue to consider a limited information scenario so that agents can only calculate their utilities but no information is available about the utilities of other users in the network. Consider a modified cooperative relaying scheme such that $\text{PU}_k$'s time slot when matched with $\text{SU}_\ell$ is now divided as follows:

\begin{itemize}
\item During the first $\frac{\tau_k(1-\tau_{k\ell})}{2}$ part of the time slot, the PU's transmitter, PT$_k$, broadcasts its data packet. The data is received by the PU's receiver, PR$_k$, and by SU's transmitter, ST$_\ell$, contingent on $h_{k\ell}\geq h_{k}$.
\item During the second $\frac{\tau_k(1-\tau_{k\ell})}{2}$ part of the time slot, ST$_\ell$ decodes the data received in phase 1 and relays $\text{PU}_k$'s message to PR$_k$ using power $P_{k\ell}$. 
\item A third time phase, ${\tau_{k\ell}\tau_k}$, is allocated by $\text{PU}_k$ for $\text{SU}_\ell$'s own communication. 
\end{itemize}

\noindent So now $\text{PU}_k$'s time slot is fixed at $\tau_k$ and a portion of that time slot is dedicated for $\text{SU}_\ell$'s communication\footnote{We will thereafter continue to assume $\tau_k=1$ w.l.o.g.}. This modification will change the agents' utilities so that
the achievable $\text{PU}_k$ rate when matched with $\text{SU}_\ell$ is 

\begin{equation}
R^{\rm P}_{k\ell}(P_{k\ell},\tau_{k\ell})=\begin{cases}
\frac{1-\tau_{k\ell}}{2}\left[R^{\rm P,dl}_{k}+\log_2\left(1+\frac{P_{k\ell}}{\sigma^2}\right)\right]&\mbox{  if  $h_{k\ell}\geq h_{k}$}\\
0&\mbox{  otherwise}.
\end{cases}
\label{rp2}
\end{equation}

\noindent The associated utility for $\text{PU}_k$ when matched with $\text{SU}_\ell$ will still be calculated as in (\ref{up}). We will also modify $\text{SU}_\ell$'s utility so that,

\begin{equation}
v_{k\ell}(P_{k\ell},\tau_{k\ell})\!\!=\!\!{\tau_{k\ell}}\log_2\left(\!\!1\!+\!\frac{\hat{P}_{k\ell}|h_{\ell}|^2}{\sigma^2}\!\!\right),
\label{us3}
\end{equation}

\noindent where $\hat{P}_{k\ell}$ is the power remaining for $\text{SU}_\ell$'s own communication after relaying for $\text{PU}_k$. We will now include a total energy constraint so that,

\begin{equation}
\frac{1-\tau_{k\ell}}{2}\frac{P_{k\ell}}{|h_{\ell k}|^2}+{\tau_{k\ell}}\hat{P}_{k\ell}=P^{\rm T}_\ell,
\label{const}
\end{equation}
\noindent where $P^{\rm T}_\ell$ is $\text{SU}_{\ell}$'s total power. We have now accounted for $\text{SU}_\ell$'s total power budget directly into its achievable rate in (\ref{us3}). We note that the secondary utility form in (\ref{us3}) is no longer quasi-concave. However $u_{k\ell}(\tau_{k\ell},P_{k\ell})$ is still decreasing in $\tau_{k\ell}$ and increasing in $P_{k\ell}$, while the opposite is true for for $v_{k\ell}(\tau_{k\ell},P_{k\ell})$\footnote{ While it is obvious $v_{k\ell}(\tau_{k\ell},P_{k\ell})$ is decreasing in $P_{k\ell}$, it is not immediately clear whether it is increasing in $\tau_{k\ell}$. However taking the derivative of $v_{k\ell}(\tau_{k\ell},P_{k\ell})$ with respect to $\tau_{k\ell}$ yields $\frac{\partial v_{k\ell}(\tau_{k\ell},P_{k\ell})}{\partial \tau_{k\ell}}=\frac{P_{k\ell}|h_{\ell}|^2/\sigma^2}{\left(\ln{2}\right)\left(1+x\right)}+\frac{1}{\ln2}\left(\ln\left(1+x\right)-\frac{x}{1+x}\right)$, where $x=\hat{P}_{k\ell}|h_{\ell}|^2/\sigma^2$. Now let $f(x)=\ln\left(1+x\right)-\frac{x}{1+x}$. Note that $x\geq 0$ in our case as it is a power term. Also note that $f(0)=0$ while $\frac{\partial f(x)}{\partial x}=\frac{x}{(1+x)^2}\geq 0$. Since $f(0)=0$ and $f(x)$ is a monotonically increasing function of $x$, we conclude that $f(x)\geq 0$, and hence $v_{k\ell}(\tau_{k\ell},P_{k\ell})$ is an increasing function of $\tau_{k\ell}$. }. Let us consider another negotiation and aspiration update process in this context.  

%%%%%%%%%%%%%%%%%%%%%%%%%%%%%%%%%%%%%%%%%%%%%%%%%%%5
\subsection{BLMA2}
%%%%%%%%%%%%%%%%%%%%%%%%%%%%%%%%%%%%%%%%%%%%%%%%%%%

Similar to the previous section, we can show that the problem of assigning PUs to SUs with the utility (\ref{up}) and the modified utility (\ref{us3}) can be formulated as a matching problem. We will now illustrate the negotiation process and aspiration update details in this context. We will collectively refer to these two procedures as BLMA2. Previously, because the utilities were quasi-convex, by connecting the line between the two agents' offers, we were assured the randomly chosen matching point provides an improvement for both agents. This is no longer true. However, we will bypass this difficulty by focusing on one dimension at any given instant while only requiring the less stringent assumption that the interests of the agents on the two sides of the market are opposed. Here are the procedures: 

\begin{enumerate}
\item \emph{BLMA2: The PU-SU negotiation process}
\begin{itemize}
\item \textbf{Initialize} $(\tau_{kl},P_{kl})$
\item \textbf{If}\,\,\,\,\,$\textsc{rand}[0,1]>\frac{1}{2}$,
\begin{enumerate} 
\item Flag$=1$.
\item Calculate $P^{\rm P}_{k\ell}$ such that $a_k+\epsilon={u_{kl}(P^{\rm P}_{k\ell}, \tau_{kl})}$,
\item Calculate $P^{\rm S}_{k\ell}$ such that $b_\ell+\epsilon={v_{kl}(P^{\rm S}_{k\ell}, \tau_{kl})}$,
\end{enumerate}
\,\,\,\, \textbf{Else}
\begin{enumerate} 
\item Flag$=0$.
\item Calculate $\tau^{\rm P}_{k\ell}$ such that $a_k+\epsilon={u_{kl}(P_{k\ell}, \tau^{\rm P}_{kl})}$,
\item Calculate $\tau^{\rm S}_{k\ell}$ such that $b_\ell+\epsilon={v_{kl}(P_{k\ell}, \tau^{\rm S}_{kl})}$,
\end{enumerate}
\textbf{End If}
\item \textbf{If}\,\,\,\,\,\\
$\left\lfloor{u_{k\ell}(\tau_{k\ell},P^{\rm S}_{k\ell})}\right\rfloor_\delta\geq a_k+\epsilon$ \textbf{and} $\left\lfloor{v_{k\ell}(\tau_{k\ell},P^{\rm P}_{k\ell})}\right\rfloor_\delta\geq b_\ell+\epsilon$\\
\textbf{OR} \\
$\left\lfloor{u_{k\ell}(\tau^{\rm S}_{k\ell},P_{k\ell})}\right\rfloor_\delta\geq a_k+\epsilon$ \textbf{and} $\left\lfloor{v_{k\ell}(\tau^{\rm P}_{k\ell},P_{k\ell})}\right\rfloor_\delta\geq b_\ell+\epsilon$\\

\,\,\,\, \textbf{Then}
\,\,\,\,\,\,\, ${\cal A}_{k\ell}(a_k+\epsilon,b_\ell+\epsilon)=1$.\\
\textbf{End If}
\end{itemize}
\item \emph{BLMA2: Aspiration Update}
\begin{itemize}
\item \textbf{If}\,\,\,\,\,Flag$=1$, \\
\,\,\,\,\,\,\,Choose $P_{kl}$ uniformly at random in $[P^{\rm P}_{k\ell},P^{\rm S}_{k\ell}]$.\\
\,\,\,\, \textbf{Else}\\
\,\,\,\,Choose $\tau_{kl}$ uniformly at random in $[\tau^{\rm S}_{k\ell},\tau^{\rm P}_{k\ell}]$.\\
\textbf{End If}
\item Update $a_k\leftarrow\left\lfloor{u_{kl}(P_{k\ell}, \tau_{kl})}\right\rfloor_\delta$,
\item Update $b_\ell\leftarrow\left\lfloor{v_{kl}(P_{k\ell}, \tau_{kl})}\right\rfloor_\delta$,
\end{itemize}
\end{enumerate}

\noindent Note that this time, users initialize with arbitrary time and power offers $(\tau_{kl},P_{kl})$. Agents then flip a coin. If the outcome is more than one half, they calculate their aspiration based power levels given the existing time. Otherwise they calculate their aspiration based time request/offer given the existing power. 

\begin{proposition}
Given $u_{k\ell}(\tau_{k\ell},P_{k\ell})$ and $v_{k\ell}(\tau_{k\ell},P_{k\ell})$ with opposing trends in $\tau_{k\ell}$ and $P_{k\ell}$, and $\epsilon=q\delta>0$ for some $\theset{q \st q>1 \text{ and } q\in {\mathbb Z}}$, BLMA2 will converge to an $\epsilon$-pairwise stable state with probability one. 
\end{proposition}

\proof

The proof proceeds as we did in Proposition 1. The only difference now is that due to the lack of quasi-convexity, we can no longer connect any line segment between agents' time and power offers, and assume that any point in between these two offers will be agreeable. We can, however, make use of the fact that users have opposing interests. Fixing time, if $\left\lfloor{u_{k\ell}(\tau_{k\ell},P^{\rm S}_{k\ell})}\right\rfloor_\delta\geq a_k+\epsilon$ {and} $\left\lfloor{v_{k\ell}(\tau_{k\ell},P^{\rm P}_{k\ell})}\right\rfloor_\delta\geq b_\ell+\epsilon$, then any point on the vertical line connecting $P^{\rm P}_{k\ell}$ and $P^{\rm S}_{k\ell}$ must be agreeable. A similar statement can be said about choosing any point on the horizontal line connecting the time offers $\tau^{\rm P}_{k\ell}$ and $\tau^{\rm S}_{k\ell}$. Fig. \ref{fig:2D2} illustrates the process of choosing the match point in BLMA2 for the sample case of the SU utility not being quasi-concave. Once agreement is established, the proof follows as in Proposition 1. 
\QED

\begin{figure}
	\centering
		\includegraphics[scale=0.6]{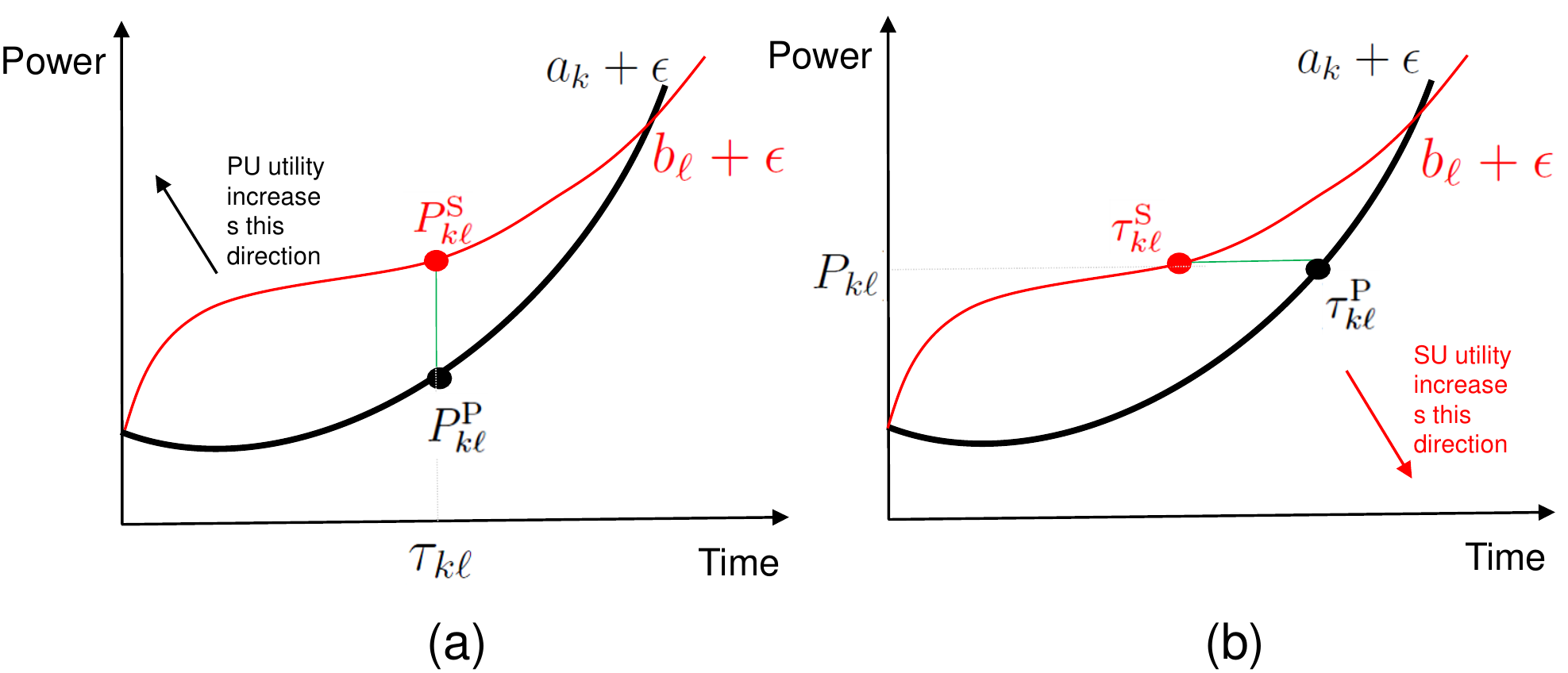}
	\caption{BLMA2: a) If $\textsc{rand}[0,1]\geq\frac{1}{2}$, $\text{PU}_k$ and $\text{SU}_\ell$ fix the time offer and search for agreements in the power offers, b) otherwise, $\text{PU}_k$ and $\text{SU}_\ell$ fix the power offer and search for agreeable matches in the time offers.}
	\label{fig:2D2}
\end{figure}

%%%%%%%%%%%%%%%%%%%%%%%%%%%%%%%%%%%%%%%%%%%%%%%%%%%%%%%%%%%%%%
\section{Numerical Results}
\label{sec:num}
%%%%%%%%%%%%%%%%%%%%%%%%%%%%%%%%%%%%%%%%%%%%%%%%%%%%%%%%%%%%%%
In this section, we evaluate the performance of the BLMA in cognitive radio networks. We randomly place the PUs and the SUs in a $1\times 1$ square area. We consider large scale fading so that the channel coefficients are computed as the inverse of the distances between the transmitting and receiving nodes. The PU transmit power $P_k=0.01$ for all PUs and $P^{\rm T}_\ell=1$ for all SUs and $\sigma^2=1$. We also take $\delta=0.05$ and $\epsilon=3\delta$.  

%%%%%%%%%%%%%%%%%%%%%%%%%%%%%%%%%%%%%%%%%%%%%%%%%%%
\subsection{Validating Proof technique}

In Fig. \ref{fig:blma1}, we plot the performance of the BLMA1 used to solve the matching problem of Section \ref{sec:cog} where we assumed quasi-convexity. In this particular figure, we plot the index of the SU matched to a particular PU and the utilities of the secondary users. Since the market is unbalanced, there are 3 PUs and 5 SUs, two SUs ($\text{SU}_1$ and $\text{SU}_2$ in this example) are left unmatched  with aspiration levels zero. This run shows a sample realization that is close to our proof technique. First, agents' aspiration levels steadily increase in the so-called pre-stable phase. This stage happens fast, in about 24 steps. Then agents compete over matches, $\text{SU}_1$ and $\text{SU}_2$ compete over $\text{PU}_3$ while $\text{SU}_4$ steadily reduces it aspiration level as it is single. $\text{SU}_4$ eventually ``wins'' at around 1400 steps. Single agents then steadily decrease their aspirations until $\epsilon$-pairwise stability is reached which happens in this realization at around 1800 steps. It is possible to achieve faster convergence if we use an adaptive $\delta$ scheme similar to the one suggested in \cite{adaptive}. %We also note that c

\begin{figure}
	\centering
		\includegraphics[scale=0.6]{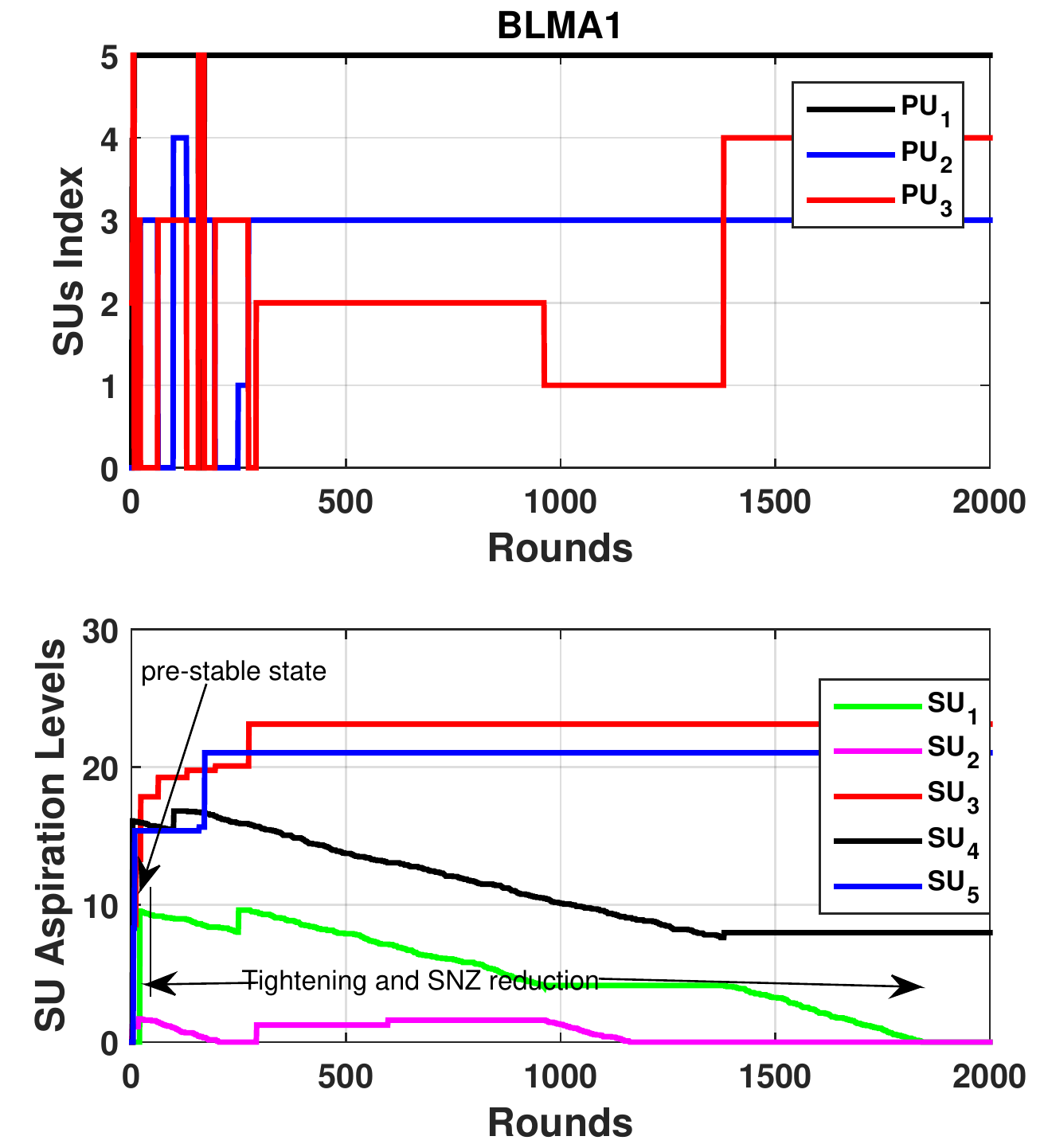}
	\caption{Number of negotiation rounds till stable matching occurs for BLMA1 in a 3 PUs $\times$ 5 SUs case.  }
	\label{fig:blma1}
\end{figure}

%%%%%%%%%%%%%%%%%%%%%%%%%%%%%%%%%%%%%%%%%%%%%
\subsection{A Complete Information Reference Scheme}

Next, we contrast our results with the scenario proposed in \cite{fengdistmat}. As mentioned earlier, the authors consider two scenarios regarding assumptions about the $\theta_{k\ell}$'s. We will contrast with the partially incomplete information scenario where PUs know the type of all SUs connected to them. This is the stronger information assumption in the work of \cite{fengdistmat}. Once this information is available, a PU can calculate the SU utility for a given time and power offer. It is in that sense that we refer to the scheme of \cite{fengdistmat} as a complete information benchmark since in our work agents know nothing about the utilities of other users in the network even if the time and power offers are known. 

Fig. \ref{fig:pu3} shows the resulting PU utility using BLMA1 and the algorithm in \cite{fengdistmat} denoted as Feng et al. The algorithm in that work resembles an ascending auction, PUs start with the time and power offers that give the least utility to SUs, and gradually increase the utility offers to SUs until the first stable allocation of time and power is reached. It is clear that the first stable matching occurring using this algorithm is the best from the PUs perspective. Since there is only one sequence in which events can happen in that algorithm, there is no need to average the algorithm results for a given channels' allocation. In our algorithm, on the other hand, the agents randomly encounter each other and so we average our results over 1000 runs of the algorithm to account for different ways in which convergence may occur. For both algorithms, we plot the resulting PUs' utilities for both cases of a balanced ($3\text{ PUs}\times3\text{ SUs}$) case and an unbalanced market  ($3\text{ PUs}\times5\text{ SUs}$). In the case of a balanced market, BLMA1 converges in an average of $1464$ rounds and converges in $5338$ rounds in the unbalanced case. Convergence takes longer in unbalanced markets since all singles' aspiration levels must be zero for stability to occur. The aglorithm in \cite{fengdistmat} converges in $1782$ rounds. 

For this simulation example, whether the market is balanced is irrelevant to the algorithm in \cite{fengdistmat} and subsequently there is no difference in the achievable utility in both cases. This is the case since agents know the types of SUs connected to them. Stability occurs before any offers can be made to ``weak'' agents. Due to lack of information about other agents, this is not the case in our algorithm and we note an appreciable increase in the PUs' utilities from the balanced to the unbalanced market case. The performance of BLMA1 in the unbalanced market case is clearly close to the perfect information scheme of \cite{fengdistmat}. If we compare with the weaker information assumption of \cite{fengdistmat}, we expect to find instances when our algorithm performs better despite the limited information restriction.

\begin{figure}
	\centering
		\includegraphics[scale=0.5]{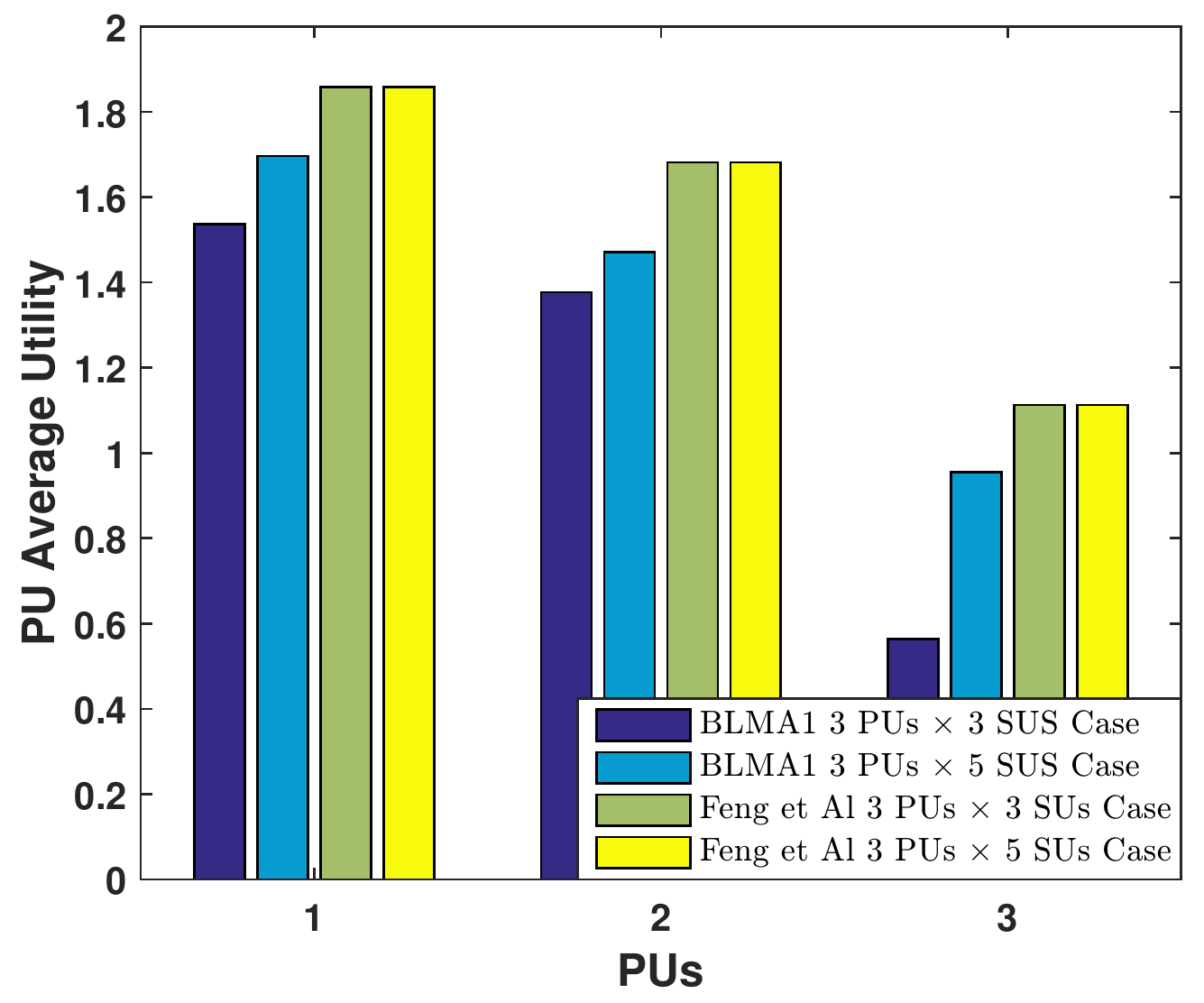}
	\caption{PUs' utility using BLMA1 and the perfect information approach in \cite{fengdistmat} for the case of balanced and unbalanced markets.  }
	\label{fig:pu3}
\end{figure}

%%%%%%%%%%%%%%%%%%%%%%%%%%%%%%%%%%%%%%%%%%%%%%%%%%%%%%%%%%%%%%
\subsection{Biasing the Market towards PU-favored Outcomes}
%%%%%%%%%%%%%%%%%%%%%%%%%%%%%%%%%%%%%%%%%%%%%%%%%%%%%%%%%%%%%%

A question remains as to the possibility of modifying the negotiation process so that the outcome favors one side of the market. As noted in \cite{roth_book}, general assignment games with {nonlinear} utilities retain the lattice structure which is characteristic of matching markets. This means that there are PU-favored outcomes, which are the least favored to the SUs, and SU-favored outcomes, which also are the least favored to the PUs. In between these two extremes are payoff vectors with varying degree of satisfaction for the users on the two sides of the market \cite{lattice}. 

In our model, a PU's utility increases with power and decreases with time so that the highest utility values occur at low time offers and large power requests. This means that the matching most favorable to all PUs will be the one where all time offers are set to zero and all power requests set to the maximum possible value per the SUs' energy constraints. This, however, would require a complete information scenario since the maximum possible power value is an SU's private information\footnote{If such information were available, then there are many techniques, as discussed in \cite{fengdistmat} for the PUs to reach a stable matching}. Furthermore, the SUs would, naturally, be reluctant to share such information given that it will yield zero secondary utility. 

We propose a simple solution: Fix all PU time offers to a specified small value. The rationale is that choosing small time offers will automatically focus the matching outcome towards PU-favored matchings. Intuitively, this proposed technique will simply be a special case of BLMA2, i.e. only search the power domain for agreements while fixing the time offers/requests to small values. However, for comparison purposes, we will call this realization 1D-BLMA.

In Fig. \ref{fig:pu1}, we compare the performance of BLMA2 and 1D-BLMA in terms of the achievable utility for PUs and SUs in both a balanced and an unbalanced market. It is clear that implementing the 1D-BLMA in both the balanced and unbalanced market cases substantially increases the utility of PUs. Furthermore, an unbalanced market with more SUs than PUs provides an advantage for PUs since it increases the competition among SUs to win matches. 

\begin{figure}
	\centering
		\includegraphics[scale=0.5]{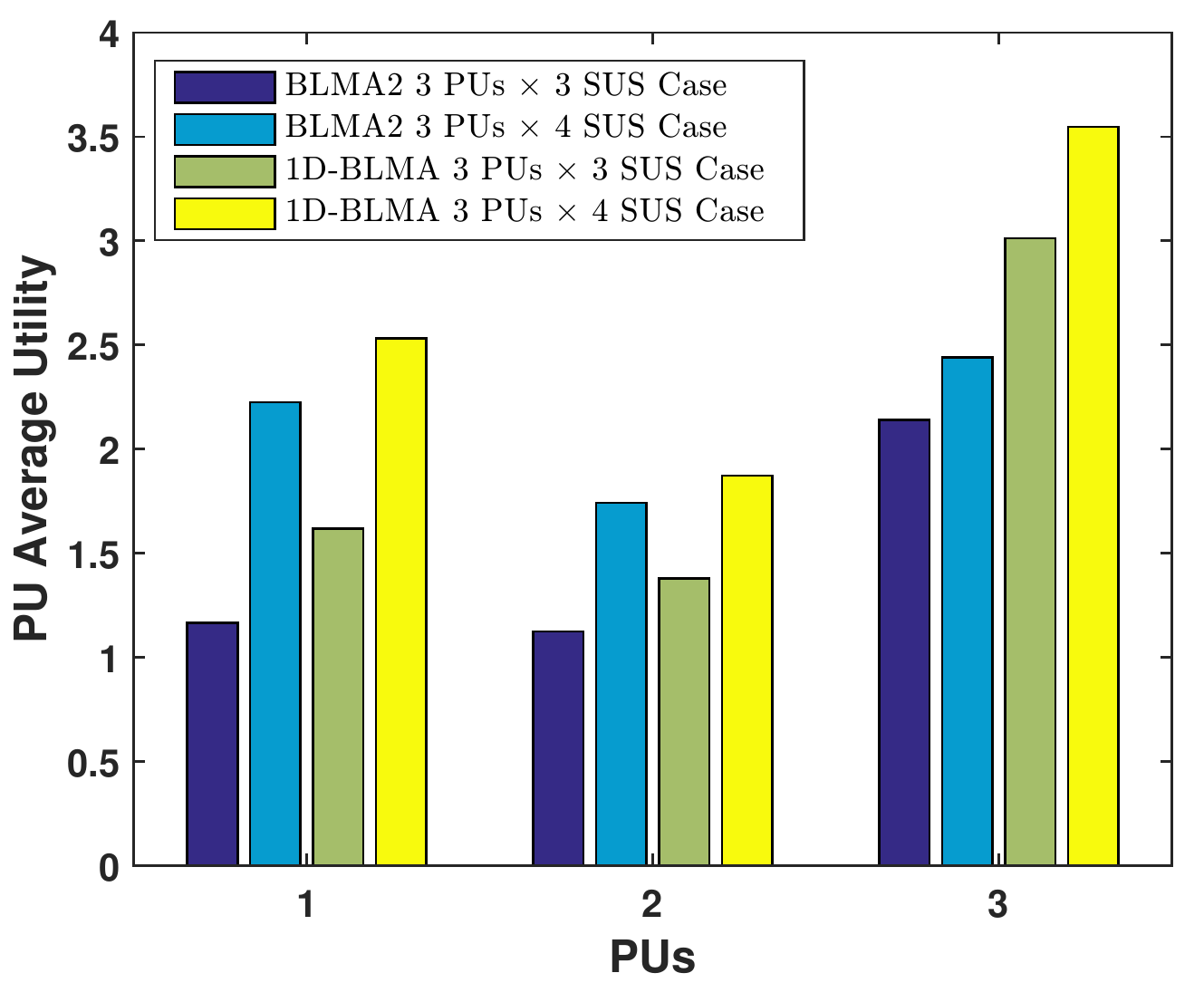}
	\caption{Average utility attained by each PU using 1D-BLMA and BLMA2 for the 3 PUs $\times$ 3 SUs case and the 3 PUs $\times$ 4 SUs case. The results are averaged over 1000 runs of the algorithms. For the 1D-BLMA, the time offers are set by all PUs to $0.1$.  }
	\label{fig:pu1}
\end{figure}

1D-BLMA also provides a faster way to implement the BLMA since the search space is reduced, now users only have to search for power values. This is evident in Table \ref{table1} which shows the average convergence times for the 1D-BLMA and BLMA2 for the case considered in Fig. \ref{fig:pu1}.

\begin{table}%[ht]
\renewcommand{\arraystretch}{1.5}
\begin{center}
\begin{tabular}{ c |l|l|l|l|l }
\hline
& \multicolumn{2}{|c|}{BLMA2}&\multicolumn{2}{|c|}{1D-BLMA}\\\hline\hline
&Balanced Market& Unbalanced Market&Balanced Market&Unbalanced Market\\\hline
Average Convergence Time (rounds)&114&315&48&85\\\hline
		
\end{tabular}
\ \\[12pt]\caption{Average convergence time for the simulation case of Fig. \ref{fig:pu1}.}
\label{table1}
\end{center}
%\vspace{-1cm}
\end{table}

%%%%%%%%%%%%%%%%%%%%%%%%%%%%%%%%%%%%%%%%%%%%%%%%%%%%%%%%%%%%%%
\section{Conclusion}
\label{sec:conc}
%%%%%%%%%%%%%%%%%%%%%%%%%%%%%%%%%%%%%%%%%%%%%%%%%%%%%%%%%%%%%%

We considered a context-free matching problem defined by agents' agreement functions. We defined a notion of pairwise stability called $\epsilon$-pairwise stability as the equilibrium concept in our model. We proposed an abstract algorithm, BLMA, to pair agents in this generic two-sided market without specifying detailed negotiation mechanisms or the actual allocation of utilities among users. The BLMA was shown to converge to $\epsilon$-pairwise stable solutions with probability one. 

Next, we considered the application of the BLMA to a cooperative relaying scheme in cognitive radio networks. Within this framework, we provided two examples of cognitive networks where users have quasi-linear utilities and more general utility forms. For the quasi-linear utility case, we showed a procedure for the BLMA, BLMA1, that exploits users' convex level sets to obtain stable solutions. For the case of non-linear utilities that still retained users' opposed interests in the optimization variables, we provided another procedure, BLMA2, to specify a negotiation mechanism and a way to update users' aspirations without the need for quasi-convexity. We also proposed a simple technique, 1D-BLMA, to bias the matching outcome towards one side of the market, the PUs' side in our case. In all such applications of the BLMA to cognitive networks, we stipulated a minimum information exchange so that users could only calculate their utilities, but no information is available about the utilities of other users in the network. Comparing with the stronger information assumptions of previous work in the literature, we showed the application of BLMA in cognitive radios can achieve close-to-optimal performance despite the information limitation. Furthermore, we note that our approach does not preclude the possibility of adding some {\emph{structure}} to the algorithm realizations, for practical implementation and for improved convergence. For example, PUs maybe activated in a round-robin fashion or the $\delta$ step reduction in aspirations can be made adaptive as the algorithm progresses. In all such case, our results will still follow as long as we can guarantee that agents have a positive probability of making a match if it exists.

In the end, the BLMA process is simple. Agents with agreeable functions, match with positive probability, otherwise, they lower their aspirations by a small step and wait for their next random activation round. We conclude that these proposed simple dynamics of BLMA can be extended to any such setting where users can be separated into two disjoint sets with opposed interests.

%%%%%%%%%%%%%%%%%%%%%%%%%%%%%%%%%%%%%%%%%%%%%%%%%%%%%%%%%%%%%

%%%%%%%%%%%%%%%%%%%%%%%%%%%%%%%%%%%%%%%%%%%%%%%%%%%%%%%%%%%%%%

%----------------------------------------------------------------------------------
\bibliographystyle{IEEEbib}
\bibliography{MyLib}
%----------------------------------------------------------------------------------

%--------------------------------  Figures and Tables  ----------------------------

%----------------------------------------------------------------------------------

\end{document}